\documentclass{aa}  
\usepackage{graphicx}
\usepackage{subfigure}
\usepackage{tabularx}
\usepackage{multirow}
\usepackage{comment}
\usepackage{natbib}
\usepackage{txfonts}
\usepackage{gensymb}
\usepackage{multirow, makecell}
\usepackage{pdflscape}
\usepackage{tablefootnote}
\usepackage{hyperref}
\hypersetup{
    colorlinks=true,
    linkcolor=blue,
    filecolor=blue,      
    urlcolor=blue,
    citecolor=blue,
}
\let\linenumbers\nolinenumbers\nolinenumbers
\begin{document}

\title{Study of SVOM/ECLAIRs inhomogeneities in the detection plane below 8~keV and their mitigation for the trigger performances}

\author{
    Xie, Wenjin\inst{1} \and
    Cordier, Bertrand\inst{1} \and
    Dagoneau, Nicolas\inst{2} \and
    Schanne, Stéphane\inst{1} \and
    Atteia, Jean-Luc\inst{3} \and
    Bouchet, Laurent\inst{3} \and
    Godet, Olivier\inst{3}
}

\institute{Université Paris Cité, CEA Paris-Saclay, IRFU/DAp-AIM, F-91191 Gif-sur-Yvette, France\\
    \email{wenjin.xie@cea.fr}
    \and
    CEA Paris-Saclay, IRFU/DEDIP, F-91191 Gif-sur-Yvette, France
    \and
    IRAP, CNRS/UPS, 9 avenue du Colonel Roche, 31028 Toulouse, France
}

\date{Received XXX; accepted XXX}

\abstract
{The Space-based multi-band astronomical Variable Objects Monitor (SVOM) is a Chinese-French mission dedicated to the study of the transient sky. It is scheduled to start operations in 2024. ECLAIRs is a coded-mask telescope with a large field of view. It is designed to detect and localize gamma-ray bursts (GRBs) in the energy range from 4 keV up to 120 keV. In 2021, the ECLAIRs telescope underwent various calibration campaigns in vacuum test-chambers to evaluate its performance. Between 4 and 8 keV, the counting response of the detection plane shows inhomogeneities between pixels from different production batches. The efficiency inhomogeneity is caused by low-efficiency pixels (LEPs) from one of the two batches, together with high-threshold pixels (HTPs) whose threshold was raised to avoid cross-talk effects. In addition, some unexpected noise was found in the detection plane regions close to the heat pipes.}
{We study the impact of these inhomogeneities and of the heat-pipe noise at low energies on the ECLAIRs onboard triggers. We propose different strategies in order to mitigate these impacts and to improve the onboard trigger performance.}
{We analyzed the data from the calibration campaigns and performed simulations with the ground model of the ECLAIRs trigger software in order to design and evaluate the different strategies.
Most of the impact of HTPs can be corrected for by excluding HTPs from the trigger processing. To correct for the impact of LEPs, an efficiency correction in the shadowgram seems to be a good solution. An effective solution for the heat-pipe noise is selecting the noisy pixels and ignoring their data in the 4--8 keV band during the data analysis.
}
{The trigger threshold is the minimum value of the signal-to-noise ratio (S/N) that is required to claim that ECLAIRs has detected a candidate event that is not related to a background fluctuation.
After introducing the efficiency inhomogeneity in the imaging simulation, the trigger threshold in the 4--8 keV band increased by a factor of 5.75 times
and 1.43 times due to the impact of HTPs and LEPs, respectively, in the worst case (on a timescale of about 20 min). The trigger threshold value was restored to its normal value after we applied an efficiency-correction method.
Introducing the heat-pipe noise in our simulations in the worst case (timescale of about 20 min) resulted in an increase in the a trigger threshold of approximately 100$\%$ in the 4--8 keV band compared to observations without heat-pipe noise. Moreover, even with this increased threshold, we estimated a false-trigger rate of 99.26$\%$ in the 4--8 keV band and 4.44$\%$ in the 4--120 keV band.
 By accepting a loss of 2.5$\%$--5$\%$ noisy pixels in the 4-8 keV energy band, we can prevent false triggers caused by heat-pipe noise and reduce the threshold increment to about 20$\%$ for the longest timescale (about 20 min) of the ECLAIRs trigger in the 4--8 keV range.
 }
{}


\maketitle

\section{Introduction} \label{introduction}


Gamma-ray bursts (GRBs) are the most cataclysmic explosive phenomena in the Universe.
Despite decades of observations and theoretical studies, many questions about GRB physics remain open \citep{zhang_2018} concerning the central engine, the jet geometry (\citealp{abbott2017gravitational}, \citealt{xie2020}), particle acceleration, and the radiation mechanism \citep{gao2015photosphere}, and about the nature of soft GRBs such as X-ray flashes (\citealp{sakamoto2008}, \citealp{bi2018}) and so on. Extending the detection energy band down to soft X-rays is important for detecting soft GRB events and enhancing our understanding of GRB physics. Additionally, it may provide some clues on the presence of new spectral features in the prompt emission of classical GRBs.
There were different wide-field X-ray cameras in the Beppo Satellite italiano per Astronomia X (BeppoSAX) \citep{boella1997} and High-Energy Transient Explorer 2 (HETE-2, \citep{ricker2003} mission, but the Wide Field Cameras (WFC, 2--30 keV) on BeppoSAX had no onboard trigger, and the HETE-2 Wide-Field X-Ray Monitor (WXM) had an onboard trigger, but its energy range was restricted to 2--25 keV.

The SVOM mission (\citealt{Wei2016}, \citealt{Cordier2015}) is a collaboration between China and France that aims to study the GRB phenomenon and the transient sky in general. The mission is scheduled to be launched in June 2024. The SVOM satellite is equipped with four scientific instruments: a soft gamma-ray telescope with a wide field of view called ECLAIRs \citep{Godet2014}, a gamma-ray spectrometer named Gamma-Ray Monitor (GRM; \citealt{wen2021}), and two narrow-field follow-up telescopes in the X-ray and visible bands. These are the Microchannel X-ray Telescope (MXT; \citealt{Gotz2023}) and the Visible Telescope (VT; \citealt{fan2020}).
In the SVOM mission, the ECLAIRs instrument provides the GRB triggers with a localization that is accurate enough to perform follow-up observations. The satellite will perform an autonomous slew onto the source for follow-up observations by MXT and VT. At the same time, these alerts will be broadcast to the observer community to allow follow-up by ground-based telescopes.
By observing GRBs across different wavelengths, the SVOM mission is expected to provide new insights into the nature of these explosive events.

ECLAIRs is a coded-mask imaging telescope with a designed low-energy threshold of 4 keV \citep{Godet2014}. It is composed of a coded mask, a detection plane \citep{Lacombe2014}, an onboard electronics called Scientific Trigger and Control Unit (UGTS; \citealt{Schanne2013}, \citealt{LeProvost2013}) implementing the GRB trigger, and a structure ensuring the geometrical rigidity, X-ray collimation, and thermal control of the instrument. The active area of the ECLAIRs detection plane is approximately 1000 cm$^{2}$, and the energy range is 4--120 keV, with a wide field of view of 2 sr (partially coded). The detection plane is divided into eight independent sectors, each composed of 5x5 modules and their readout electronics. 
Each module is made of an array of $4 \times 8$ ($4\times 4$ mm$^2$ and 1 mm thick) Schottky-type CdTe detectors from Acrorad Ltd (Japan) hybridized with the low-noise and low-consumption ASIC IDef-X from CEA (\citealt{2009ITNS...56.2351G}), totaling 6400 CdTe detectors \citep{Remoue2010}. This enables an energy threshold at 4 keV.

The combination of a coded mask with a detection plane sensitive from 4 keV 
enables the ECLAIRs telescope to achieve a competitive sensitivity despite the modest volume and mass allocation imposed by the platform.
With this low energy threshold, ECLAIRs will be notably sensitive to soft GRBs and X-ray rich transients \citep{Godet_2009}.


Two different triggers are embedded in the UGTS, namely the count-rate trigger (CRT) and the image trigger (IMT).
The CRT performs short-duration observations between 10 ms and 20 s, while the IMT is used for longer observations that last between 20.48 s and about 20 min. For the CRT, the detection plane is divided into nine different zones (four quadrants, four halves, and the full detector) and four different energy bands (e.g., 4--8 keV, 8--20 keV, 8--50 keV, and 8--120 keV). Then the number of detected counts is compared with the predicted background count rate in different energy bands and detector zones for different timescales. If the detected counts exceed a threshold for a given energy band, detector zone, and timescale, a sky-image reconstruction is performed by deconvolution of the detection plane image (called shadowgram) by the mask pattern \citep{goldwurm2003}. Then, new transient sources are searched for in the sky images.

The IMT (\citealt{Dagoneau2022}) subtracts the background in 20.48 s shadowgrams, reconstructs the corresponding sky image by deconvolution, and stacks multiple sky images up to 1310.72 s. New transient sources are systematically searched for in all sky images. Background subtraction, deconvolution, and transient search are performed in four configurable energy bands (e.g., 4--8 keV, 8--20 keV, 8--50 keV, and 8--120 keV).

Both trigger algorithms are executed in parallel in the UGTS.
The different steps of the CRT and IMT can be deeply configured with many parameters. 
In this study, we focus on the count-shadowgram preprocessing and the following deconvolution to build the sky images. 
To adjust the contribution of each pixel in the trigger algorithm, three parameters per energy band are used in the onboard trigger software. They are the efficiency value, weight value, and efficiency limit.

\begin{itemize}
\item The efficiency value is the efficiency of the pixel. The detected count will be corrected for this efficiency value.
\item The weight value is the pixel weight in the background subtraction and deconvolution process.
\item The efficiency limit is a scalar value that is the criterion for deciding whether the pixel is used. When the efficiency of a pixel is below this value, this pixel is ignored in the trigger process.
\end{itemize}

From the efficiency value and weight value, two tables are built: the efficiency tables, and the weight tables (one of each per energy band).
Each of these two tables can be selected by configuration for use in the following operations: for the pixel efficiency correction, for the pixel contribution for the background fit, and for the pixel contribution for the deconvolution (per energy band).
Typically, the efficiency table is used for the pixel efficiency correction prior to the deconvolution. The contribution of each pixel for the background fit (IMT only) and the deconvolution is tuned by the weight table.

The efficiency correction is performed by dividing the detected counts by the efficiency to obtain the corrected shadowgram in counts and by the square of the efficiency to obtain the variance of the shadowgram. If the efficiency is 0, both counts and variances are set to 0. The count and the variance corrections can be activated or deactivated separately in the configuration. If the variance normalization is deactivated, the shadowgram in counts (possibly corrected by the efficiency) is also used as the variance.
The way to set the pixel contribution for the deconvolution is presented in \cite{goldwurm2003}.

The onboard software updates the initial values for the two tables. It sets the values to 0 for which the initial efficiency values are lower than or equal to the efficiency limits. As an example, with these configuration parameters (but with smaller matrices),

\begin{align}
\text{efficiency configuration} &= \begin{bmatrix}
1 & 0.2\\
0.5 & 0.9\\
\end{bmatrix};\\
\text{weight configuration} &= \begin{bmatrix}
1.0 & 1.0\\
1.0 & 1.0\\
\end{bmatrix};\\
\text{efficiency limit} &= 0.5.
\end{align}

The two tables used by the onboard software will be

\begin{align}
\text{efficiency} &= \begin{bmatrix}
1 & 0\\
0 & 0.9\\
\end{bmatrix};\\
\text{weight} &= \begin{bmatrix}
1.0 & 0\\
0 & 1.0\\
\end{bmatrix}.
\end{align}


In 2021, a series of test campaigns were performed on the ECLAIRs flight model in a thermal vacuum test-chamber (TVAC) at the French Space Agency (CNES) in Toulouse. This model will be integrated in the SVOM satellite. 
The main goal of these campaigns was to study the detailed performances of the ECLAIRs detection plane and to set up the camera parameters (in particular, the low-energy threshold values for the 6400 pixels).
The experimental sequences, setup, ground segment equipment, main performances, and calibration results are detailed in \citealt{Godet2022}. The TVAC tests of ECLAIRs used different sources of X-ray photons: beams of X-ray fluorescence photons produced with an X-ray generator radiating on metal targets (lines between 4--22 keV, depending on the target material), and calibrated radioactive sources.

We analyzed the datasets of the 2021 campaigns. Intriguingly, we found some effects that degraded the performances in the 4--8 keV band. These effects included some inhomogeneities in the detector pixel efficiency, and an additional noise in two regions of the detection plane close to the heat pipes. Due to this spatial proximity, we call this the heat-pipe noise, even though the link between this noise and the heat pipes remains unproven. In \cite{Godet2022}, this was called the structured low-energy (SLE) noise excess. 
In Sect. \ref{Sec:experiment} we present the details of these inhomogeneities in the ECLAIRs detection plane in the 4--8 keV band that were observed during the TVAC test. A detailed investigation of their impact on the ECLAIRs trigger performances, as well as some possible solutions for mitigating their effects, is reported in Sect. \ref{Sec:inhomogeneity}. The impact of the heat-pipe noise and methods for mitigating its influence are detailed in Sect. \ref{Sec:Heat-pipe noise}. A discussion and conclusion are presented in Sect. \ref{Sec:Discussion} and \ref{Sec:Conclusion}.

\section{Experiment carried out in 2021, and the data analysis in the 4--8 keV band } \label{Sec:experiment}

\begin{figure*}[!h]
    \centering
    \includegraphics[width=0.8\textwidth]{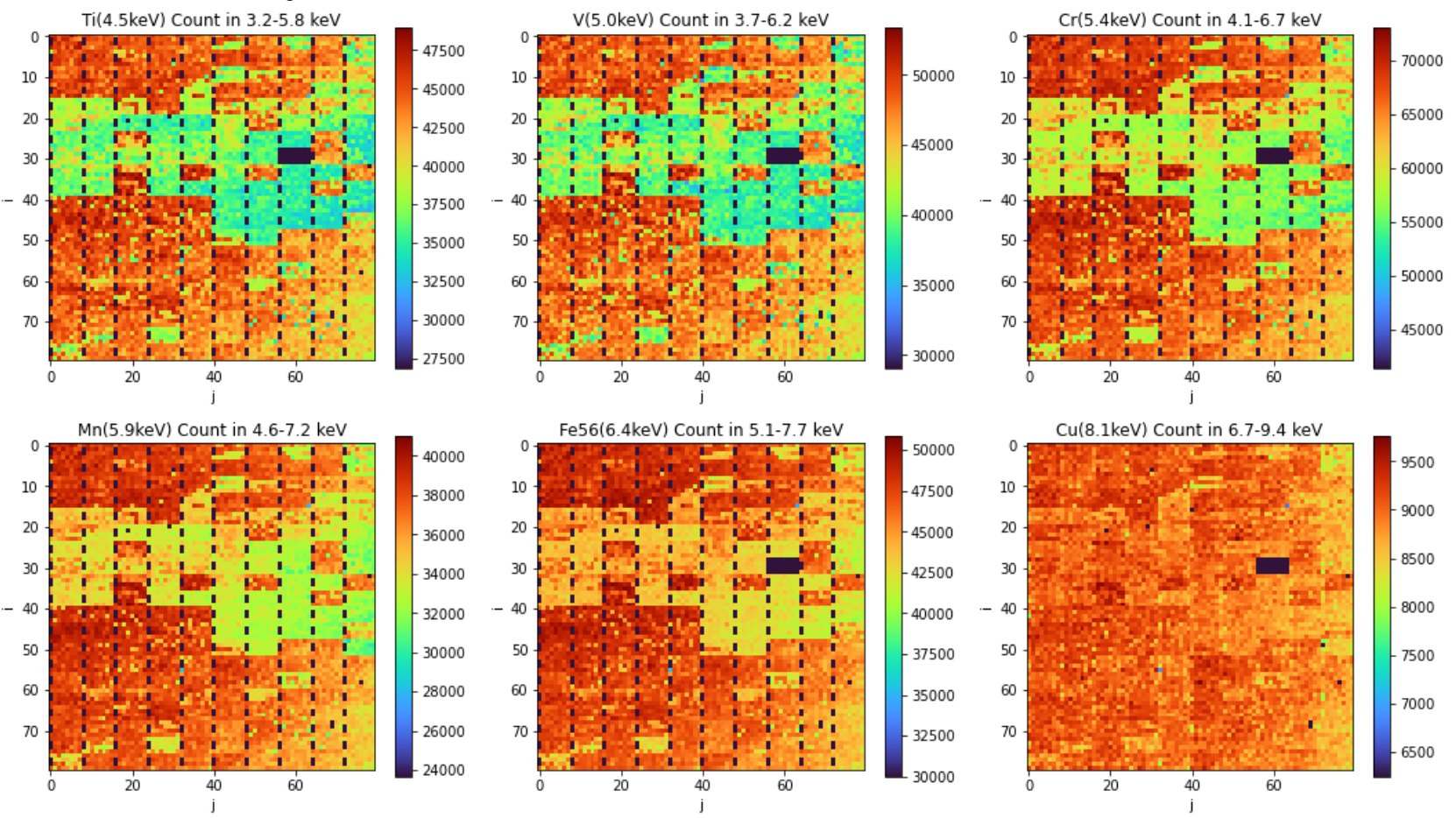}
    \caption{Images of the detected number of counts per pixel in the detection plane between 4.5 and 8 keV, obtained with six X-ray generator targets (producing X-ray fluorescence lines of a given energy): $^{22}$Ti (4.5 keV), $^{23}$V (5.0 keV), $^{24}$Cr (5.4 keV), $^{25}$Mn (5.9 keV), $^{26}$Fe (6.4 keV), and $^{29}$Cu (8.0 keV). The energy bands are determined to be $\pm 3\sigma$ around the mean value of the best Gaussian fit for the spectrum of the whole detector plane. (One detector module of 4×8 pixels in the middle right does not work nominally, except for the $^{25}$Mn target.) }
    \label{fig:fig1}
\end{figure*}

\begin{figure*}[!h]
    \centering
    \includegraphics[width=1.0\textwidth]{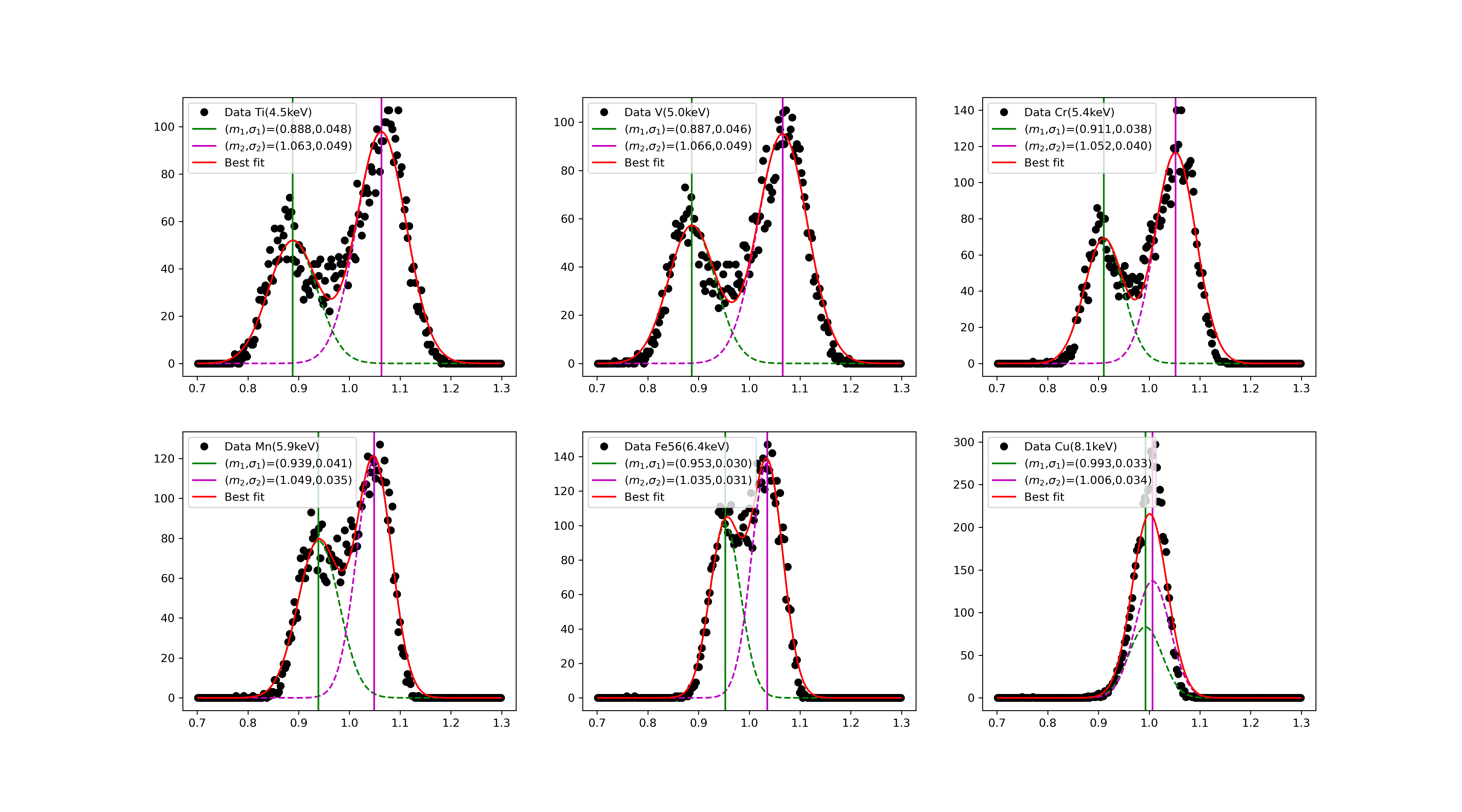}
    \caption{Histograms of the relative number of counts in the six images (corresponding to the six energy bands) shown in Fig. \ref{fig:fig1}. The two populations of LEPs and HEPs appear. The mean number of relative counts is determined by Gaussian fits. The HTPs are removed from the histograms.}
    \label{fig:fig2}
\end{figure*}

In this section, we present the performances of the detection plane in the 4--8 keV band. 
We analyzed the datasets obtained from the TVAC calibration campaigns. We used data when the detection plane was illuminated with a beam of X-ray fluorescence photons. The X-ray beams produced by an X-ray generator plus various targets are
$^{22}$Ti (4.5 keV fluorescence line), $^{23}$V (5.0 keV), $^{24}$Cr (5.4 keV), $^{25}$Mn (5.9 keV), $^{26}$Fe (6.4 keV), and $^{29}$Cu (8.0 keV).

\subsection{Efficiency inhomogeneity below 8 keV}

The count distributions in the detection plane for different targets are shown in Fig. \ref{fig:fig1} for photon energies ranging from 4 keV to 8 keV. 
The considered energy intervals for each target were taken to be $\pm 3\sigma$ around the Gaussian peak when we fit the spectrum of the counts detected in the whole detection plane. 
The distribution displays counting inhomogeneities below 8 keV, and part of the pixels have even lower counts. 
The counting inhomogeneity of the X-ray illumination during the calibration is about 2 \% (std/mean). This was estimated in a dedicated Geant4 simulation, and this value is consistent with the inhomogeneity measured with the copper target (2.9 \%, std/mean).

Based on the difference in counting efficiency, we identify three pixel populations. The first population consists of the higher-efficiency pixels (HEPs). They accounts for the majority of the pixels (about two-thirds). The second population consists of the relatively lower-efficiency pixels (LEPs), which represent approximately one-third of the total pixels. The third population named high-threshold pixels (HTPs) includes the dashed black lines of pixels shown in Fig \ref{fig:fig1}. 

The HTPs are caused by a cross-talk effect between the tracks within the 32-pixel elementary module \citep{Godet2022}. The cross-talk is connected with the routing of the tracks in the analog and digital boards of the module. The analog tracks of pixels 8 and 16 (which typically transport micro-volt signals) pass very close to the track transporting the trigger signal on the digital board (which typically transports volt signals). As a consequence, when one of the 32 pixels of a detection module triggers, the trigger signal induces a small signal in the analog tracks of pixels 8 and 16, which causes these pixels to trigger simultaneously.

Since the analog signal in pixels 8 and 16 is small, it is not visible above 7 keV, and the solution adopted to remove this cross-talk was to raise the energy threshold of pixels 8 and 16 to a level equivalent to an energy of 7 keV.  Pixels 8 and 16 are located on the left side of each module. This explains the alignment of the HTPs in the detection plane.

For the HEPs and LEPs, the CdTe pixels of ECLAIRs were manufactured by Acrorad Ltd. from two different batches produced in 2008 and 2016. Most of the pixels with fewer counts come from the 2016 batch. This may be due to the difference in the manufacturing processes between the two batches, resulting in different detection parameters.
To quantify the number of HEP and LEP pixels and the efficiency discrepancy change with energy between them, we constructed the distributions of the relative counts between pixels, which were normalized by the mean value of the total count per pixel, showing the two populations HEP and LEP. The distributions were then fit with two Gaussian functions to model HEPs and LEPs, and the results are shown in Fig. \ref{fig:fig2}. 

Fig. \ref{fig:fig2} shows that the counting discrepancy between the two batches of detectors decreases when the energy increases.
At 8 keV, the two batches of pixels show a homogeneous efficiency distribution and cannot be distinguished. 

\begin{figure*}[!h]
\centering    
\subfigure[]{\label{fig: triggerRate a}\includegraphics[width=.4\textwidth]{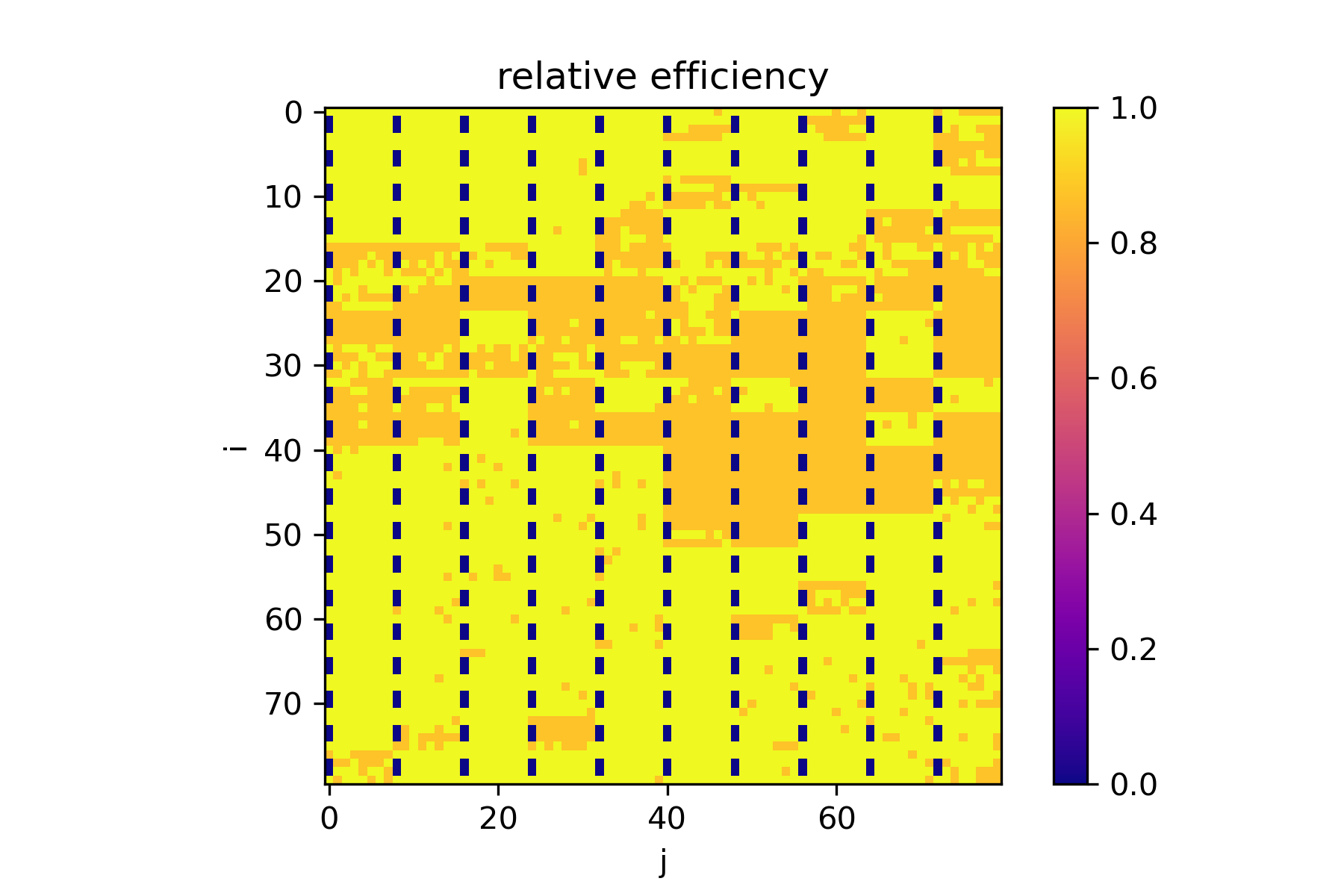}}
\subfigure[]{\label{fig: triggerRate b}\includegraphics[width=.45\textwidth,height=.28\textwidth]{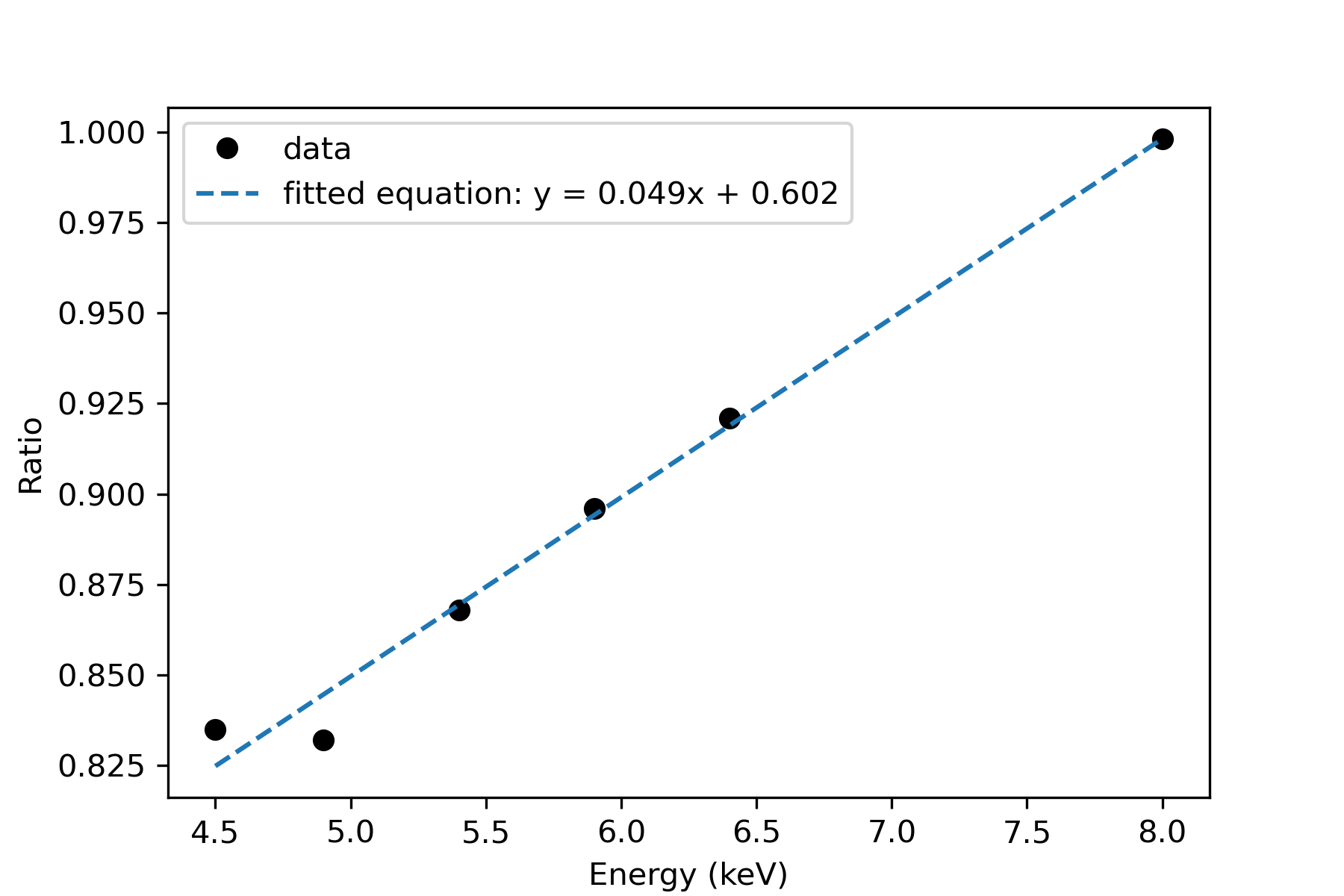}}
\caption{Three pixel populations found in the ECLAIRs detection plane (left: HTPs in black, LEPs in orange, and HEPs in yellow). The ratio of the mean number of relative counts between the LEP and the HEP as a function of energy is shown on the right, and it is fit with a linear function.}
\label{fig: Three pixel populations}
\end{figure*}

We chose the dataset of the Cr target (5.4 keV) to associate each pixel of the detection plane with one of the three populations because it is easier to separate the two Gaussian peaks, as shown in Fig.\ref{fig: Three pixel populations} (left panel). The ratio of the Gaussian mean value between the HEP and LEP populations in the 4--8 keV band is also shown in the right panel of Fig.\ref{fig: Three pixel populations}. The relation between the efficiency ratio of the LEP compared to the HEP population versus energy was fit with a simple linear equation.

\subsection{Heat-pipe noise in the 4--8 keV band}

We report abnormal noise in 4--8 keV nearby the heat pipes (heat-pipe noise) in the TVAC tests. Even though the origin of this noise is still unknown, it seems to be related to the way the heat pipes work.
We focused on data that were collected when no X-ray photons illuminated the detection plane. 
To do this, we divided the TVAC data ($2.4 \times 10^{5}$ s) into 20 s time bins and created detector count images of the entire detection plane for each time bin.
Then, we only selected the images with a total number of counts between 100 and 1500, which excluded the periods when the shutter of the X-ray source was open and periods when ECLAIRs was not in operation mode. Finally, 8690 images totaling an exposure time of $\sim$ 1.74 $\times$ 10$^{5}$ s were selected from the entire dataset.

\begin{figure*}[!h]
    \centering
     \subfigure{\label{fig:fig4a}\includegraphics[width=1.0\textwidth]{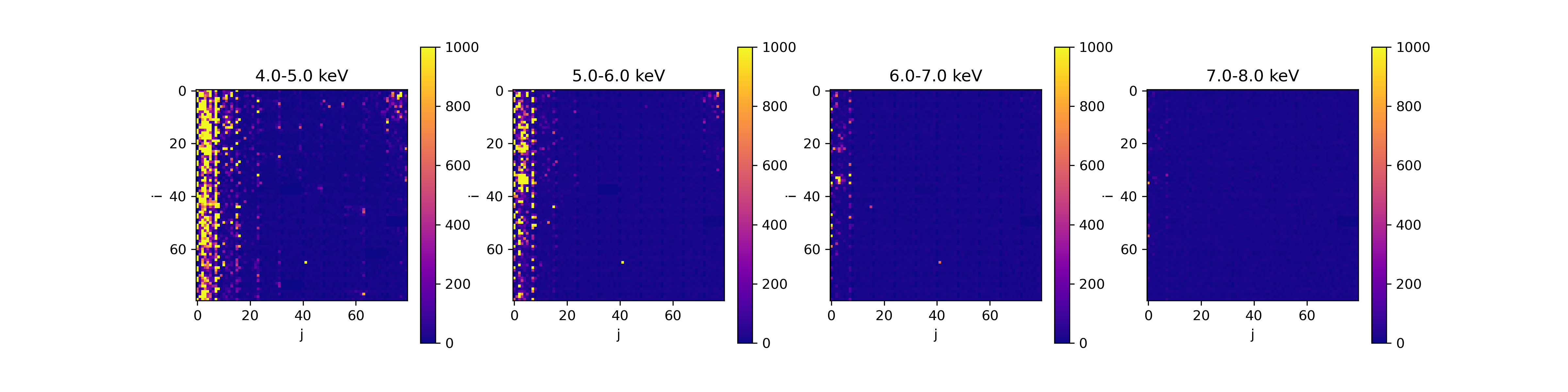}}
     \subfigure{\label{fig:fig4b}\includegraphics[width=1.0\textwidth]{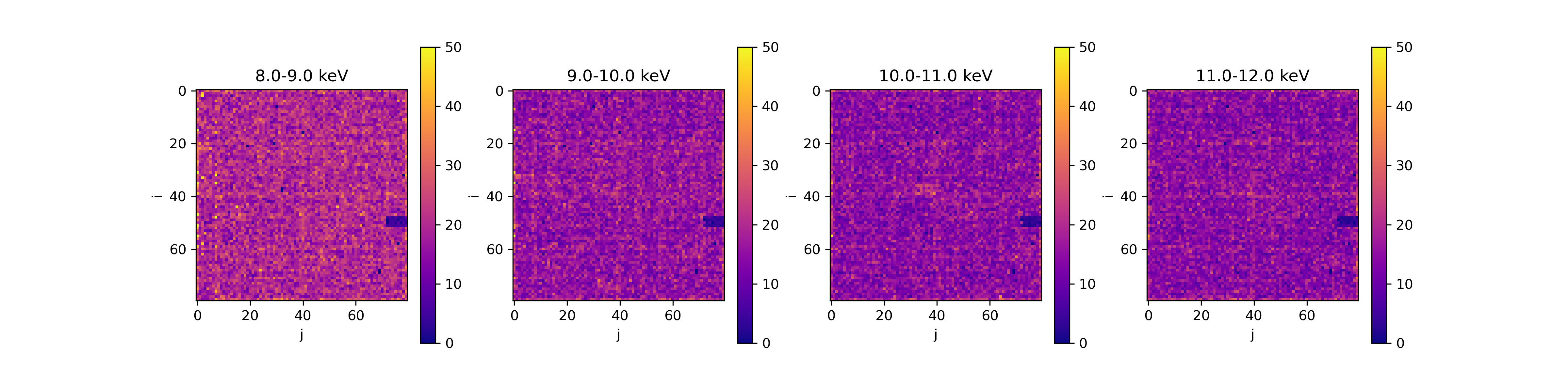}}
    \caption{Images of the detected number of counts per pixel in eight energy bands (from 4 to 12 keV) integrated over all 1.74 $\times$ 10$^{5}$ s of TVAC data where no X-ray source is present (background exposures only). The heat-pipe noise appears in the 4 to 8 keV energy band as pixels with abnormally high count rates. (One detector module of 4$\times$8 pixels on the right does not work nominally. It shows zero counts).}
    \label{fig: energy distribution of entire TVAC data }
\end{figure*}

We divided the selected dataset into eight energy bins between 4 and 15 keV and created detector-plane images for each bin and for the whole duration. As shown in Fig.\ref{fig: energy distribution of entire TVAC data }, the noise is clearly visible in the 4--8 keV energy range, which is located at the vertical edges and in the top right corner of the detection plane.
Within the 4--8 keV band, the noise count increases when the energy decreases. 
Above 8 keV, the count distribution in the detection plane becomes uniform, and the noise is negligible.

\begin{figure*}
    \centering
    \includegraphics[width=1.0\textwidth]{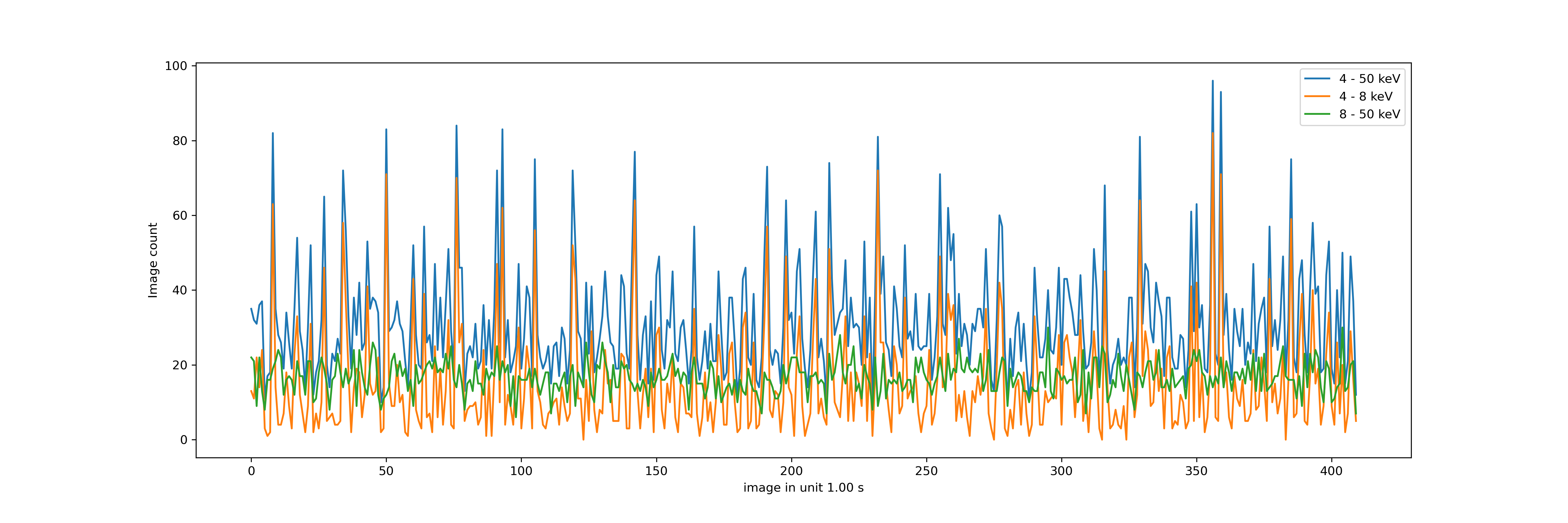}
    \caption{Light curve of the detection plane image in 1.0 s time bin during the TVAC test (background exposures only) for different energy bands: 4--8 keV (orange), 8--50 keV (green), and 4--50 keV (blue). The mean count rate in 4--8 keV is 15.2 (count/s). The background count rate in 8-50 keV is relatively stable, with a mean value of 15.8 count/s.}
    \label{fig:fig5}
\end{figure*}

The heat-pipe noise is not stable, but varies strongly with time. The light curve of the detection plane in different energy bands is shown in Fig.\ref{fig:fig5}. In the energy range of 8--50 keV, the background count rate exhibits a stable distribution with a mean value of 15.8 counts/s. However, due to the presence of heat-pipe noise within the 4--8 keV energy band, the light-curve count distribution appears peculiar and lacks regularity. The mean count rate in 4--8 keV is measured to be 15.2 counts/s, with a heat-pipe noise that could reach about five times the mean value. 

\section{Efficiency inhomogeneity} \label{Sec:inhomogeneity}

\subsection{ Impacts on the ECLAIRs trigger }

To study the impact of LEPs and HTPs on the imaging performances and on the detection of sources, we built a dedicated simulation process. We detail the steps of this process below. 

\begin{enumerate}
    \item We simulated the orbital background seen by ECLAIRs due to the CXB \citep{Moretti2009} assuming perfect performances for the detection plane in the 4--8 keV band. The simulation was processed with a Python library wrapping the same C++ code as is used for the onboard IMT (image trigger). Sixty-four detection shadowgrams were simulated, each exposure lasting 20.48 s. These shadowgrams were deconvolved, and the 64 sky images were stacked together to reach a total exposure time of 1310.72 s. In this simulation, all pixels were assumed to have an efficiency of 1.

    \item We conducted the same CXB simulation as in step 1, introducing LEP or HTP pixels. In this case, the HEP pixels were assumed to have an efficiency of 1, and the efficiency of LEP pixels was set to a value in each 1 keV bin, which depends on the ratio shown in Fig. \ref{fig: Three pixel populations}. The HTP pixel efficiency was set to 0. 
\end{enumerate}

The trigger threshold means the minimum value of the signal-to-noise ratio (S/N) that is required to claim that ECLAIRs has detected a candidate event unrelated to a background fluctuation. 
In an ideal situation, in which the shadowgram is filled with a flat background observed for a sufficiently long time, the sky S/N produced by the deconvolution follows a normal distribution with $\sigma_{S/N}$ = 1. In this case, the canonical 3$\sigma$ trigger threshold for one pixel is equivalent to 5.4$\sigma$ for 6400 pixels \citep{dagoneau:tel-03009638}. This is the solution of the product of 6400 cumulative distribution functions for the normal distribution ($\mathcal{N}(0,1)$) to reach a probability of 3$\sigma$ (ie. 99.865$\%$). This is only an ideal value that is valid under the assumption that the background counts follow a Gaussian distribution, are perfectly subtracted, and that pixels of the sky are independent. In practice, the background is not flat, even after its subtraction. The real threshold can be configured (even after launch), and its initial value was set to 6.5. This is based on a simulation in which the background was subtracted and the deconvolved sky images were analyzed with a threshold that was updated until the limit of one false alert per day was reached. Previous studies have set the trigger threshold to 6.5 $\times$ $\sigma_{S/N}$ \citep{Dagoneau2022} in order to dynamically adapt the threshold according to the S/N distribution. 

\begin{figure}[!h]
    \centering
    \includegraphics[width=0.5\textwidth]{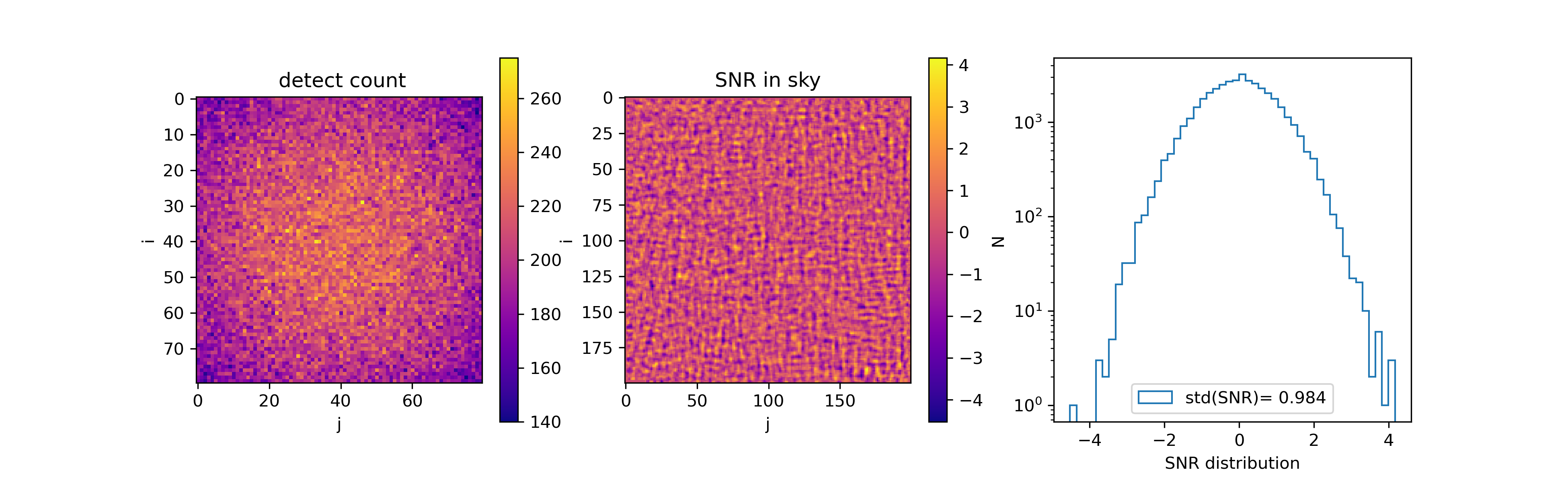}
    \includegraphics[width=0.5\textwidth]{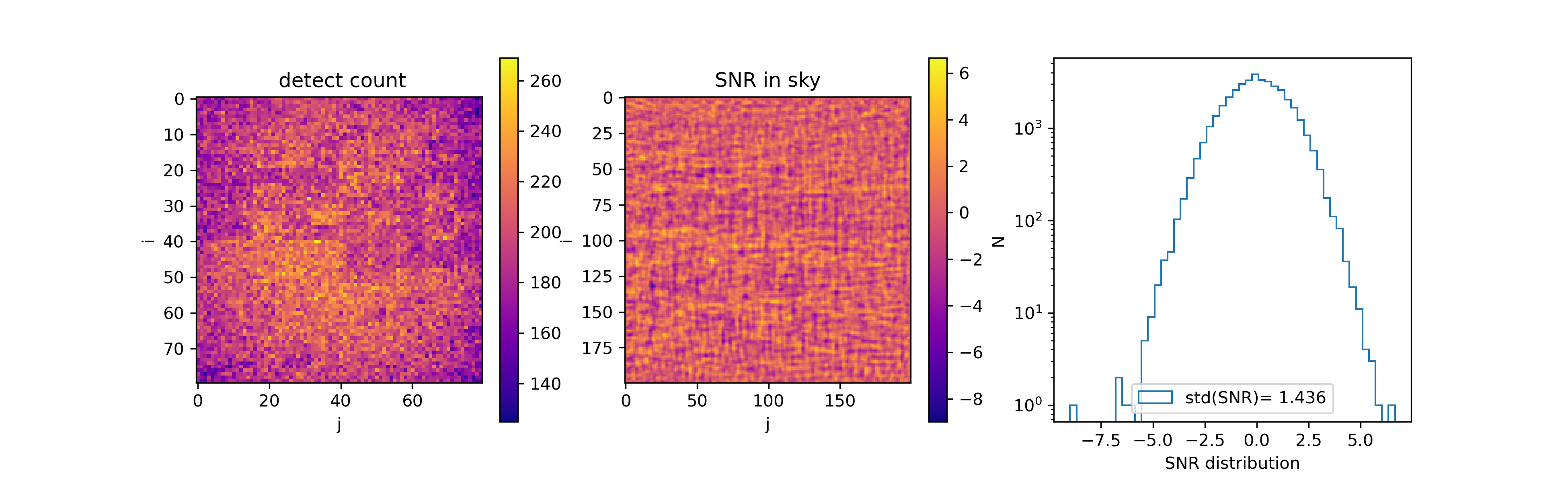}
    \includegraphics[width=0.5\textwidth]{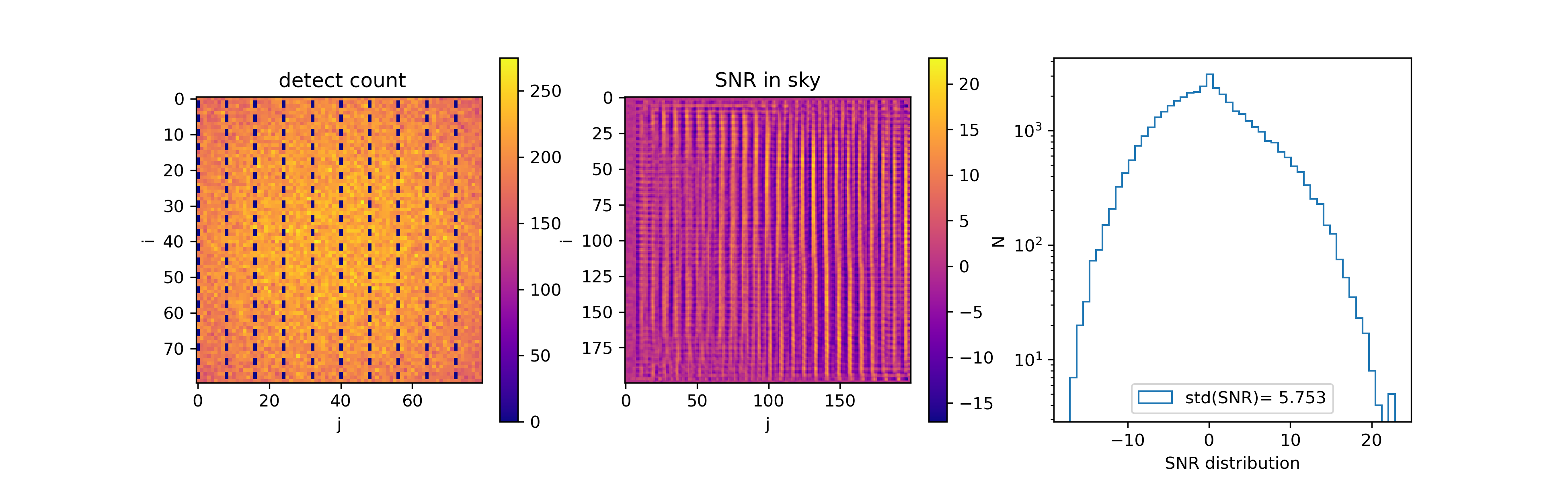}
    \caption{Simulation of the CXB on a perfect detector (row 1), with simulated LEP attenuation (row 2), and taking HTPs into account in the simulation (row 3). Left: Simulated detector-count distributions. Middle: Deconvolved sky images in S/N without any correction. Right: Distribution of the S/N of all pixels for the images shown in the middle column.}
    \label{fig: inhomegeneityExample}
\end{figure}

The simulation results for a perfect detection plane that is only exposed to the CXB, as well as the impact of LEPs and HTPs, are shown in Fig. \ref{fig: inhomegeneityExample}. In the case of the perfect detection plane (row 1), a normal S/N distribution is obtained after the deconvolution, and $\sigma_{S/N}$ is equal 0.984, which is close to the ideal theoretical value of 1. In row 2, when the LEP attenuation effect is included in the simulation, the $\sigma_{S/N}$ value of the sky image is equal to 1.436. This means that the trigger threshold in 4-8 keV would increase by 43.6$\%$ due to the LEP impact. When we consider the HTPs alone, some stripes appear in the sky image, as shown in row 3. These stripes strongly reduce the quality of the sky image, and the threshold increases by a factor of 5.753 since $\sigma_{S/N}$ = 5.753).

\subsection{Mitigation method}

\begin{figure}[!h]
    \centering
    \includegraphics[width=0.5\textwidth]{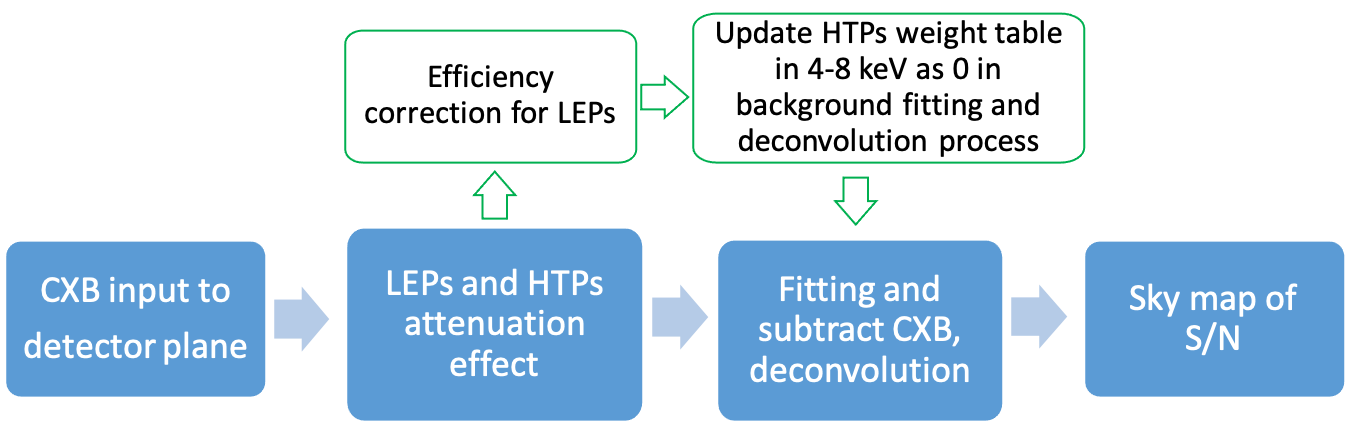}
    \caption{Simulation process after involving the mitigation methods for LEPs and HTPs. The blue blocks represent the standard imaging process without correcting for the inhomogeneity effects. The upper white blocks represent the additional actions for mitigating the impact of the LEP and HTP.}
    \label{fig: solution of LEP and HTP}
\end{figure}

To mitigate the impact of the attenuation caused by LEPs and HTPs, different solutions have been developed and applied in the processing, as shown in Fig. \ref{fig: solution of LEP and HTP}. For HTPs, the solution involves setting the weight of these pixels to 0 for the background-fitting table and for the deconvolution table in the trigger algorithm (described in Sec. \ref{introduction}). This means that the HTP pixel counts are ignored in the trigger algorithm in the 4--8 keV energy band. For LEPs, since the attenuation effect is taken into account, an efficiency correction based on the background photon spectrum was applied in order to estimate the detection counts without the LEP attenuation, which is the shadowgram subsequently used in the deconvolution process.

The trigger algorithm works in four energy bands that can be configured (Sec. \ref{introduction}). Therefore, the key is to determine the correction factor that is to be applied in each specific energy band. This correction factor depends on the spectrum of the observed source and on the efficiency of the pixels computed in this band. To compute this efficiency, we chose 1 keV bins in which the efficiency can be considered as constant. We calculated this correction factor by using the following formula:

\begin{equation}
f_{E_\text{low},E_\text{high}}=\frac{\sum_{i} \varepsilon_{i} N_{i}}{\sum_{i} N_{i}}
\label{equation1}
\end{equation}

\begin{equation}
N_{i}=\int_{i_\text{low}}^{i_\text{high}} N(E)dE,
\label{equation2}
\end{equation}

where $f_{E_\text{low},E_\text{high}}$ in \autoref{equation1} is the correction factor in the specified energy band, $i$ is the index of the 1 keV energy bin, and $\varepsilon_i$ represents the efficiency of the pixel in the bin $i$. $N_i$ stands for the source counts in bin $i$. In \autoref{equation2},  $i_{low}$ and $i_{high}$ indicate the low and high boundary of the energy bin $i$. $N(E)$ represents the CXB spectrum \citep{Moretti2009}. We chose the CXB spectrum because the CXB counts are dominant compared to the point-like sources, and because the homogeneity of the detection plane has to be primarily ensured during periods without GRBs (most of the time) in order to avoid false alerts. Moreover, in the case of a weak GRB, the background is dominant. In the case of a strong GRB, a slight inhomogeneity does not impact its detection and localization. As a result, the average efficiency correction factor $f_{4,8}$ we obtained is 0.875 in the 4--8 keV band. 

\begin{figure}[!h]
    \centering
    \includegraphics[width=0.5\textwidth]{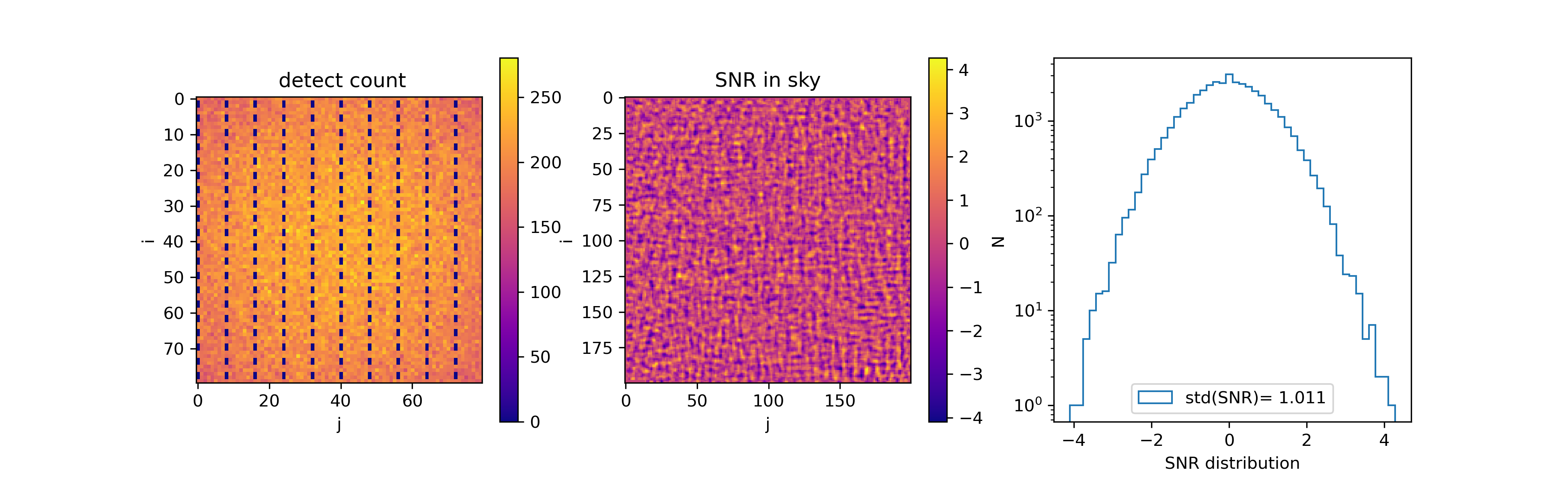}
    \caption{Simulation result in the 4--8 keV band for a 20 min exposure after applying the efficiency correction on the detector-plane image for LEPs and ignoring the counts from HTPs during the deconvolution process.}
    \label{fig: solveInhomegeneityExample}
\end{figure}

After the methods discussed above were applied in the simulation, the result of the sky S/N distribution is shown in Fig. \ref{fig: solveInhomegeneityExample}. The distribution of the sky S/N becomes uniform, and the $\sigma_{S/N}$ value decreases to 1.012, which is close to the theoretical value of the ideal model 1. Therefore, the method we used is sufficient to mitigate the impact caused by the efficiency inhomogeneity in the detection plane.

In order to demonstrate the impact of the efficiency inhomogeneity more clearly, we conducted another simulation by setting the background value equal to ten times the value of the CXB. The results are shown in Table \ref{tab:inhomogeneity simulation}.
We found that the solution we proposed effectively mitigated the impact induced by HTPs and LEPs, even in a situation where the background was equal to ten times the CXB counts. After applying the correction method, the $\sigma_{S/N}$ decreased from 16.13 to approximately 1.13.

\begin{table}[!h]
    \centering
    \caption{Simulation result of sky $\sigma_{S/N}$, which involved the LEP and HTP impact (without correction) and after applying the mitigation methods (with correction) for 20 minutes. "10 $\times$ CXB" indicates that the background value of the CXB was set to ten times the true value.}
    \begin{tabular}{|p{1.6cm}|p{2cm}|p{2cm}|p{1.6cm}|}
    \hline Background & Include effect & $\sigma_{S/N}$ without correction & $\sigma_{S/N}$ after correction \\
    \hline  { CXB } & LEP & $1.436$ & $0.997$ \\
    \cline { 2 - 4 } & HTP & $5.753$ & $0.997$ \\
    \cline { 2 - 4 } & LEP and HTP & $5.785$ & $1.011$ \\
    \hline {$10~\times$ CXB} & LEP & $3.128$ & $1.138$ \\
    \cline { 2 - 4 } & HTP & $16.059$ & $1.035$ \\
    \cline { 2 - 4 } & LEP and HTP & $16.135$ & $1.138$ \\
    \hline
    \end{tabular}
    \label{tab:inhomogeneity simulation}
\end{table}

\section{Heat-pipe noise} \label{Sec:Heat-pipe noise}

\subsection{Simulation}

In this section, we describe the simulations using TVAC data to evaluate the potential impact of the heat-pipe noise on the detection of sources with ECLAIRs. Our approach consisted of simulating the CXB and adding data extracted from TVAC tests, featuring the heat-pipe noise signature. The simulation process involved the following steps:

\begin{enumerate}
    \item We simulated the CXB and projected the counts onto the detection plane in the energy range of 4--8 keV and 4--120 keV.
    \item We added the counts from the TVAC data in the energy range of 4--8 keV to the detection plane with the same duration as the CXB. Here, the heat-pipe noise was not stable over the time, but fluctuated as shown in Fig. \ref{fig:fig5}.
    \item We processed the combined data using either the count-rate trigger (CRT, 10 ms -- 20.48 s) or the image-trigger (IMT, 20.48 s -- 1310.72 s) algorithms over different timescales.  We saved the data of the sky S/N, maximum S/N.
    \item We repeated steps 2-3 until all TVAC data were used.
\end{enumerate}

\begin{figure}[!h]
   \centering
   \includegraphics[width=0.48\textwidth]{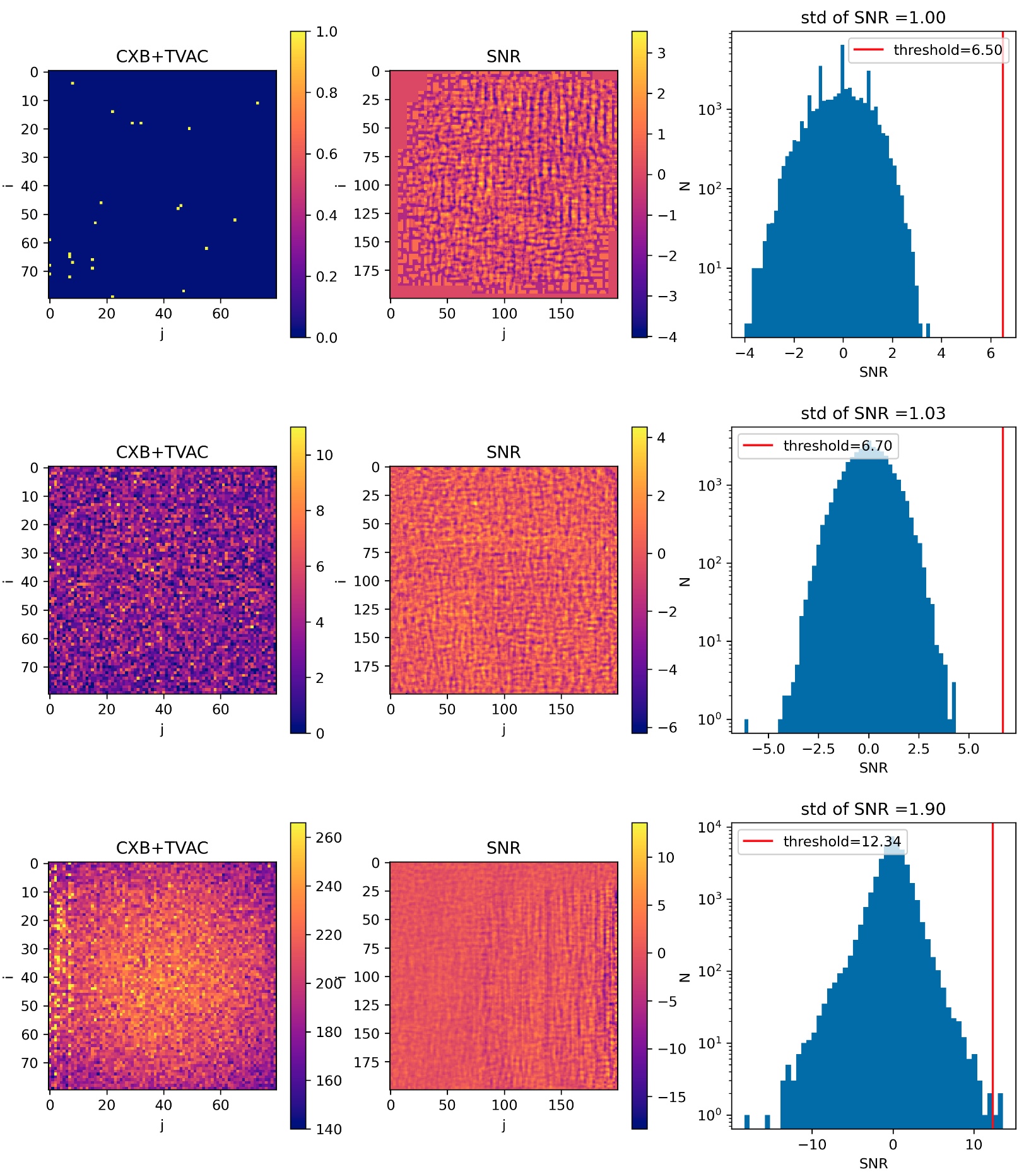}
\caption{Simulation of the heat-pipe noise impact on ECLAIRs observations for different time scales in the 4--8 keV band (without correction for heat-pipe noise). From top to bottom: 10 ms, 20.48 s, and 1310.72 s observation simulations. The panels from left to right show the CXB plus TVAC count images, S/N sky maps, and histograms of S/N sky maps. The red line in the rightmost image is the trigger threshold, the trigger threshold of ECLAIRs equal to 6.5 $\sigma_{S/N}$ (where $\sigma_{S/N}$ is the standard deviation of each S/N histogram).}
\label{fig:exampleOfImpactOfColumnEffect}
\end{figure}

Fig. \ref{fig:exampleOfImpactOfColumnEffect} presents a simulation example to assess the impact of the heat-pipe noise of ECLAIRs on 10 ms, 20.48 s, and 20 min observations in the 4--8 keV band. In the normal case (without heat-pipe noise), the $\sigma_{S/N}$ value is close to 1.0. However, when this noise component is included, the $\sigma_{S/N}$ distribution widens to 1.90 in 20 min. This means that in this case, the trigger threshold value would be increased by a factor 1.90. Nevertheless, some S/N points in the sky image still exceed the threshold, indicating that the heat-pipe noise not only increases the trigger threshold, but can also lead to additional false triggers during onboard operations.

\subsection{Impact}

\begin{figure}[!h]
    \centering
    \subfigure[]{\label{fig:stdSNR_heat-pipe noise impact a}\includegraphics[width=0.4\textwidth]{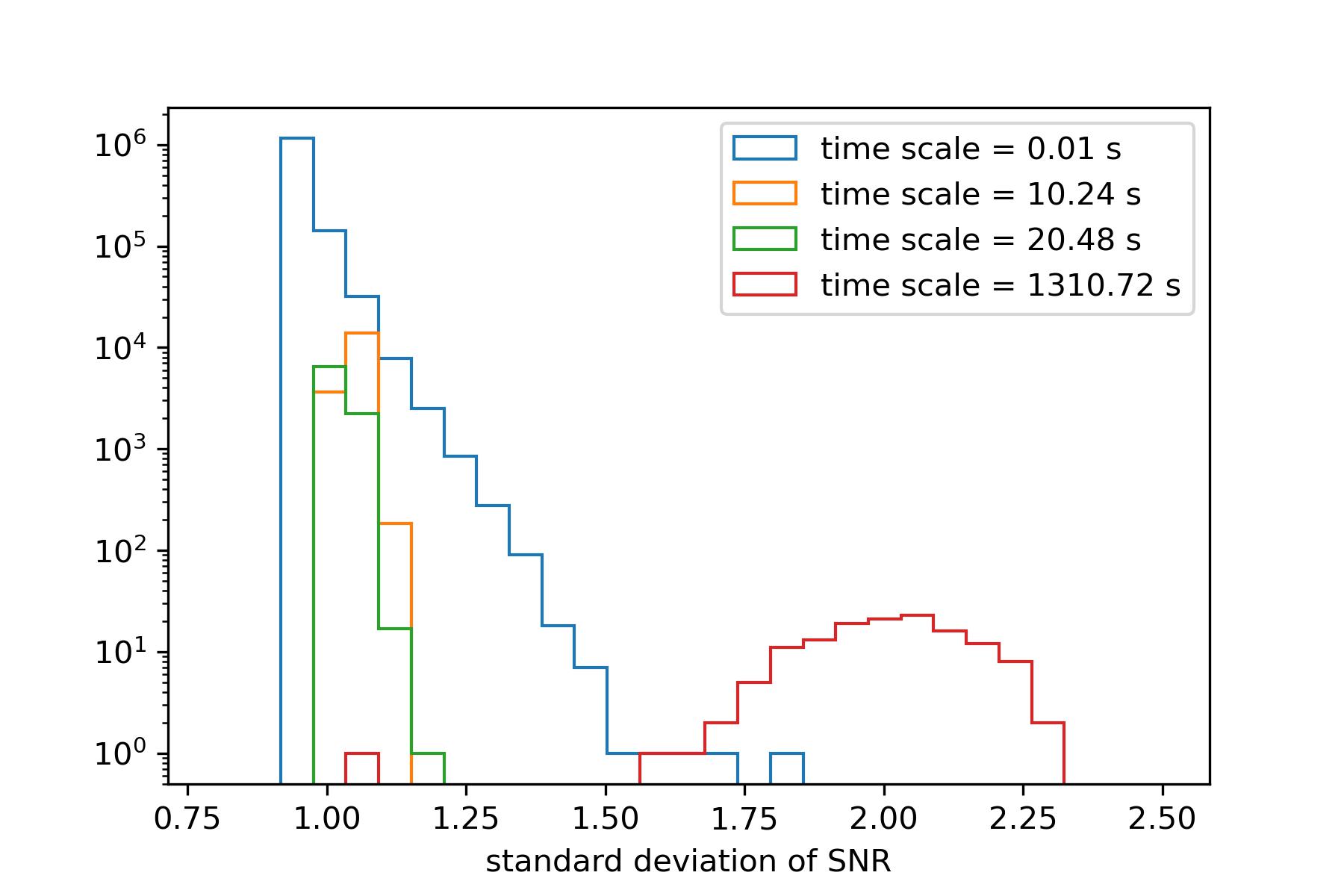}}
    \subfigure[]{\label{fig:stdSNR_heat-pipe noise impact b}\includegraphics[width=0.4\textwidth]{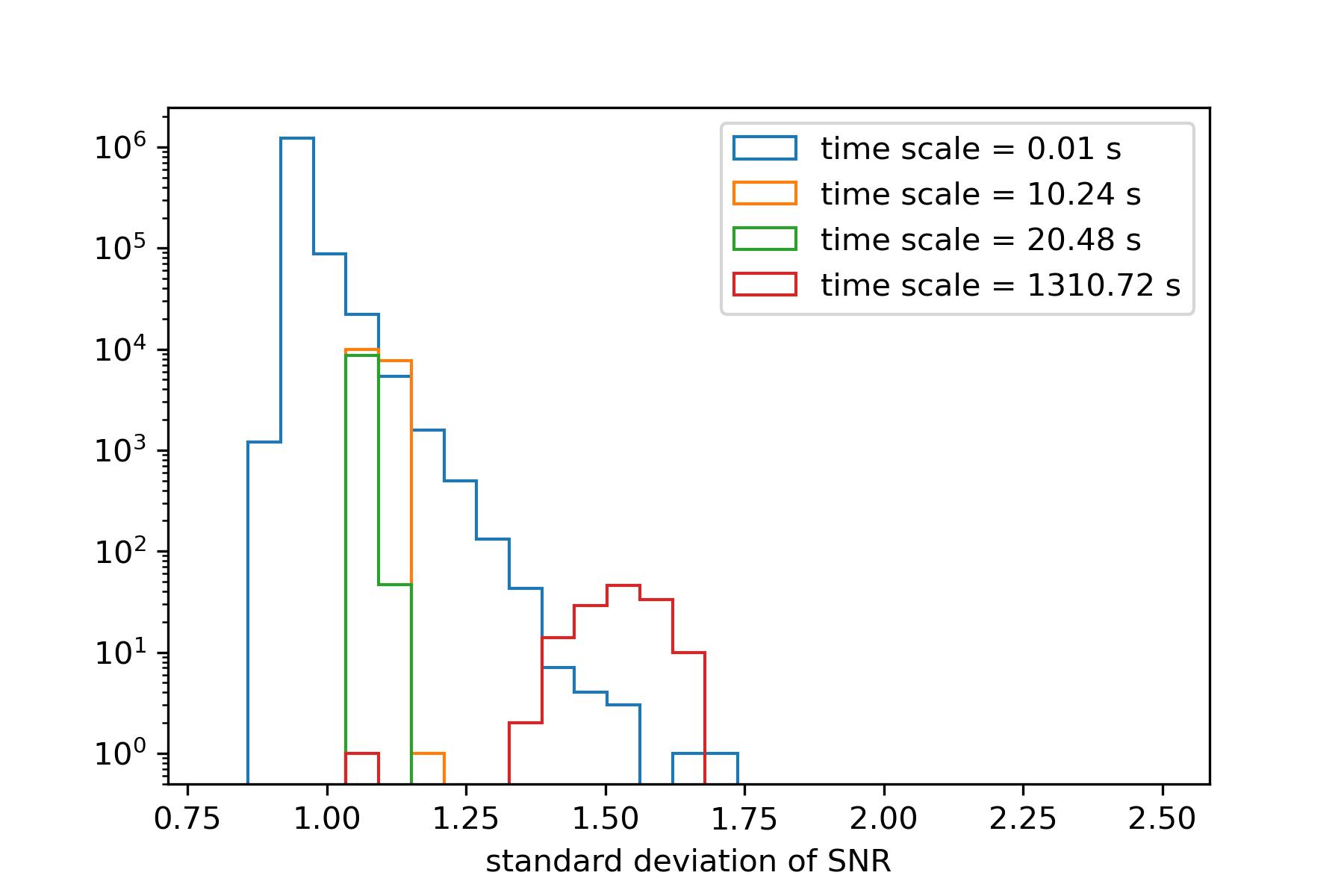}}
    \caption{Histograms of $\sigma_{S/N}$ of all deconvolved sky maps (repeating the procedure of Fig. \ref{fig:exampleOfImpactOfColumnEffect}) obtained with all TVAC data (background-only periods without an X-ray source) overlaid with simulated CXB counts in the 4--8 keV band (a) and in the 4-120 keV band (b) for different timescales of 0.01 s, 10.24 s, 20.48 s, and 1310.72 s. We divided the TVAC data into intervals with a duration that was fixed to the timescales described above, among which we selected only intervals with at least one count (removing zero-count images mostly for the short-duration timescales). The resulting numbers of data points in the histograms are 1340000, 17660, 8745, and 135 for the timescales we used.}
    \label{fig:stdSNR_2pixel}
\end{figure}

We made a simulation using all the 1.74$\times$10$^5$ s selected TVAC data for different observation timescales, ranging from 10 ms up to 1310.72 s. We found that the main impact of the heat-pipe noise on the trigger threshold mainly occurs at long timescales. For a timescale of 1310.72 s, the trigger threshold may increase by a factor of approximately 100 $\%$ in the 4--8 keV band and by 55 $\%$ in 4--120 keV band, as shown in Fig. \ref{fig:stdSNR_2pixel}. 

False triggers can be caused when the maximum sky S/N point exceeds the trigger threshold in the absence of point-like sources. We used all selected TVAC data to simulate the false-trigger rate caused by the heat-pipe noise for both the 4--8 keV and 4--120 keV bands for all timescales. The results are presented in Tables \ref{Tab:False trigger rate_CNT} and \ref{Tab:False trigger rate_IMT}. For a given timescale, we calculated the false-trigger rate as the percentage of time intervals (time slices) containing a trigger. In the 4--8 keV band, when the timescale is 20 min, the false-trigger rate (IMT only) is equal to 99.26$\%$, indicating that almost all cases experience false triggers when heat-pipe noise is present. In the 4--120 keV band, the false-trigger rate is lower than in the 4--8 keV band because the CXB photons in the 8-120 keV band help to smooth out the impact of heat-pipe noise. However, a false-trigger rate of 4.44 $\%$ in the 4--120 keV band is still caused by the accumulation of heat-pipe noise during 20 min, which corresponds to 1.6 false triggers per day. Even though we raised the trigger threshold in the simulation, meaning that true weak GRBs would not be detected, we still experience false triggers.

\begin{table*}[!h]
    \centering
    \caption{False-trigger rate resulting from heat-pipe noise in the count-rate trigger (CRT), the result from a simulation combining the simulated CXB with 1.74$\times$10$^5$ s of TVAC data. The two lines indicate the fraction of intervals showing a trigger (top) and the numbers from which this fraction is calculated (number of intervals with a trigger divided by number of independent intervals during TVAC).}
\begin{tabular}[width=14 cm]{|c|c|c|c|c|c|}
\hline Timescale (s) & 0.01 & 0.02 & 0.04 & 0.08 & 0.16 \\
\hline $\mathrm{4-8\ keV}$ & $\begin{array}{c}0 \% \\
(0 / 1340000)\end{array}$ & $\begin{array}{c}0 \% \\
(0 / 1280000)\end{array}$ & $\begin{array}{c}0 \% \\
(0 / 1190000)\end{array}$ & $\begin{array}{c}0 \% \\
(0 / 1040000)\end{array}$ & $\begin{array}{c}0 \% \\
(0 / 824375)\end{array}$ \\
\hline $\mathrm{4-120\ keV}$ & $\begin{array}{c}0 \% \\
(0 / 1340000)\end{array}$ & $\begin{array}{c}0 \% \\
(0 / 1280000)\end{array}$ & $\begin{array}{c}0 \% \\
(0 / 1190000)\end{array}$ & $\begin{array}{c}0 \% \\
(0 / 1040000)\end{array}$ & $\begin{array}{c}0 \% \\
(0 / 824375)\end{array}$ \\
\hline Timescale (s) & 0.64 & 1.28 & 2.56 & 5.12 & 10.24 \\
\hline $\mathrm{4-8\ keV}$ & $\begin{array}{c}0 \% \\
(0 / 327188)\end{array}$ & $\begin{array}{c}0.001 \% \\
(2 / 160686)\end{array}$ & $\begin{array}{c}0 \% \\
(0 / 750400)\end{array}$ & $\begin{array}{c}0.006 \% \\
(2 / 36040)\end{array}$ & $\begin{array}{c}0.136 \% \\
(24 / 17670)\end{array}$ \\
\hline $\mathrm{4-120\ keV}$ & $\begin{array}{c}0 \% \\
0\end{array}$ & $\begin{array}{c}0.002 \% \\
(3 / 160686)\end{array}$ & $\begin{array}{c}0 \% \\
(0 / 750400)\end{array}$ & $\begin{array}{c}0.006 \% \\
(5 / 36040)\end{array}$ & $\begin{array}{c}0 \% \\
(0 / 17670)\end{array}$ \\
\hline
\end{tabular}
\label{Tab:False trigger rate_CNT}
\end{table*}

\begin{table*}[!h]
    \centering
    \caption{False-trigger rate resulting from the heat-pipe noise in the image trigger (IMT), the result from a simulation combining the simulated CXB with 1.74$\times$10$^5$ s of TVAC data. The two lines indicate the fraction of intervals showing a trigger (top) and the numbers from which this fraction is calculated (number of intervals with a trigger divided by number of independent intervals during TVAC).}
\begin{tabular}{|c|c|c|c|c|c|c|c|}
\hline $\begin{array}{c}\text { Timescale } \\
\text { (s) }\end{array}$ & 20.48 & 40.96 & 81.92 & 163.84 & 327.68 & 655.36 & 1310.72 \\
\hline $\mathrm{4-8\ keV}$ & $\begin{array}{c}0.091 \% \\
(8 / 8750)\end{array}$ & $\begin{array}{c}0.069 \% \\
(3 / 4352)\end{array}$ & $\begin{array}{c}0.092 \% \\
(2 / 2170)\end{array}$ & $\begin{array}{c}0 \% \\
(0 / 1084)\end{array}$ & $\begin{array}{c}2.399 \% \\
(13 / 542)\end{array}$ & $\begin{array}{c}60.886 \% \\
(165 / 271)\end{array}$ & $\begin{array}{c}99.259 \% \\
(134 / 135)\end{array}$ \\
\hline $4-120\ \mathrm{keV}$ & $\begin{array}{c}0.126 \% \\
(11 / 8750)\end{array}$ & $\begin{array}{c}0 \% \\
(0 / 4352)\end{array}$ & $\begin{array}{c}0 \% \\
(0 / 2170)\end{array}$ & $\begin{array}{c}0 \% \\
(0 / 1084)\end{array}$ & $\begin{array}{c}0 \% \\
(0 / 542)\end{array}$ & $\begin{array}{c}0 \% \\
(0 / 271)\end{array}$ & $\begin{array}{c}4.444 \% \\
(6 / 135)\end{array}$ \\
\hline
\end{tabular}
\label{Tab:False trigger rate_IMT}
\end{table*}

\subsection{Mitigation methods} \label{subsection: Heat-pipe noise mitigation methods}

An effective solution to correct for the heat-pipe noise is to select the noisy pixels and ignore their data in the 4--8 keV band. This is realized by setting the weight of the selected noisy pixels to 0 in the background fit table and in the deconvolution table (as described in Sec. \ref{introduction}). We identified two methods for selecting the noisy pixels based on the count frequency in 10 ms time bins: the frequency selection, and the distribution selection.

\begin{enumerate}
    \item Frequency selection. For a given pixel, we counted the number of 10 ms time slices ($N_{freq}$) in which the pixel detected at least one photon, regardless of the number of detected photons. We set a threshold to identify noisy pixels as pixels having a high number of $N_{freq}$. For example, we removed pixels for which $N_{freq}$ was larger than 4200 in the 1340000 intervals, as shown in Fig. \ref{fig:frequency_selection}. This selection depends on only one parameter, $N_{freq}$.

\begin{figure}[!h]
    \centering
    \includegraphics[width=0.49\textwidth]{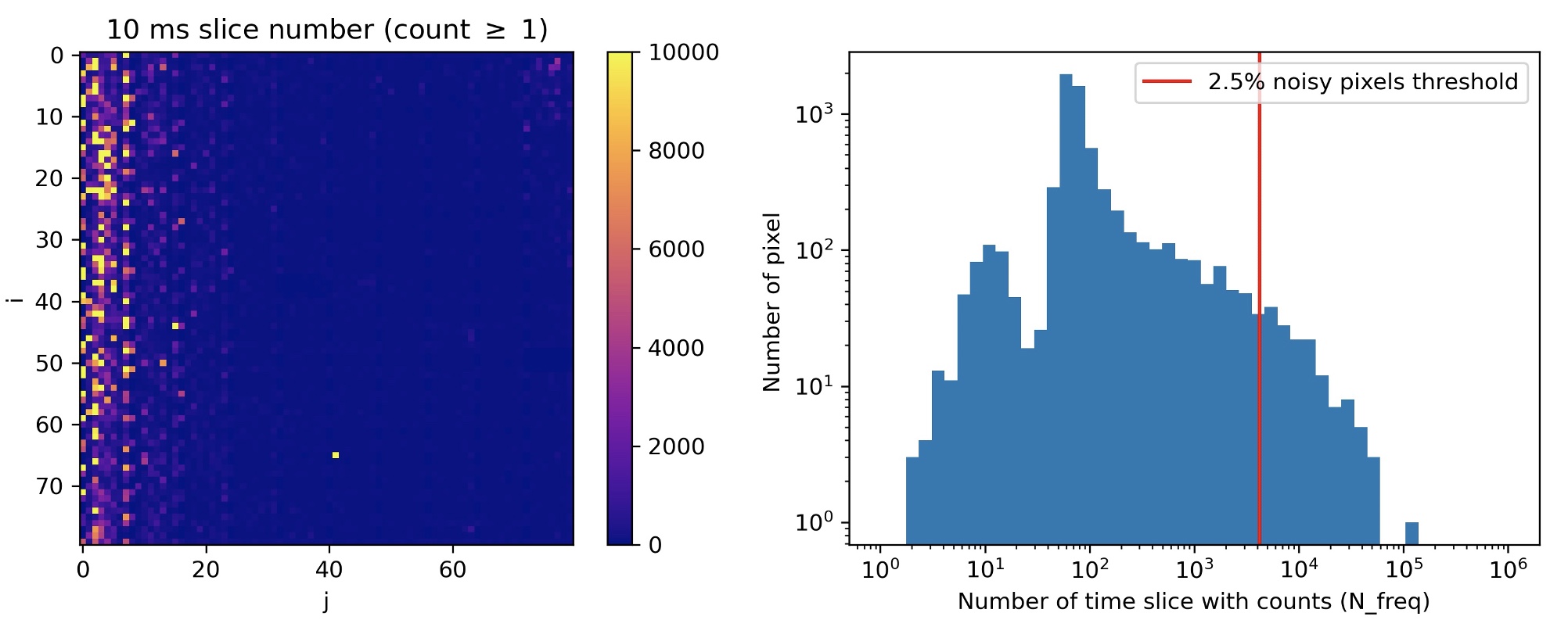}
    \caption{Frequency selection method. Left: 10 ms slice number ($N_{freq}$) distribution of pixels with a count $\ge$ 1 in the detection plane. Right: Histogram of $N_{freq}$ for 6400 pixels corresponding to the left figure. The red line is the threshold of $N_{freq}$ for selecting 2.5\% noisy pixels.}
    \label{fig:frequency_selection}
\end{figure}

    \item Distribution selection. For each pixel, we counted the number of photons in each 10 ms time slice ($N_{photon}$). Then, we counted the number of time slices corresponding to individual $N_{photon}$ to build an integral distribution of $N_{freq\_p}$. 
    To identify noisy pixels, we first chose a minimum value for $N_{photon}$ and then set a threshold on $N_{freq\_p}$. The noisy pixels have a value for the chosen bin that is higher thand the threshold. For example, we remove all pixels counting $N_{photon}$ $\ge$ 3 in 10 ms with $N_{freq\_p}$ $\ge$ 4 from our data set. This selection depends on two parameters: $N_{photon}$, and $N_{freq\_p}$. A selection example of three pixels is shown in Fig. \ref{fig:distribution_selection}.

\begin{figure}[!h]
    \centering
    \includegraphics[width=0.3\textwidth]{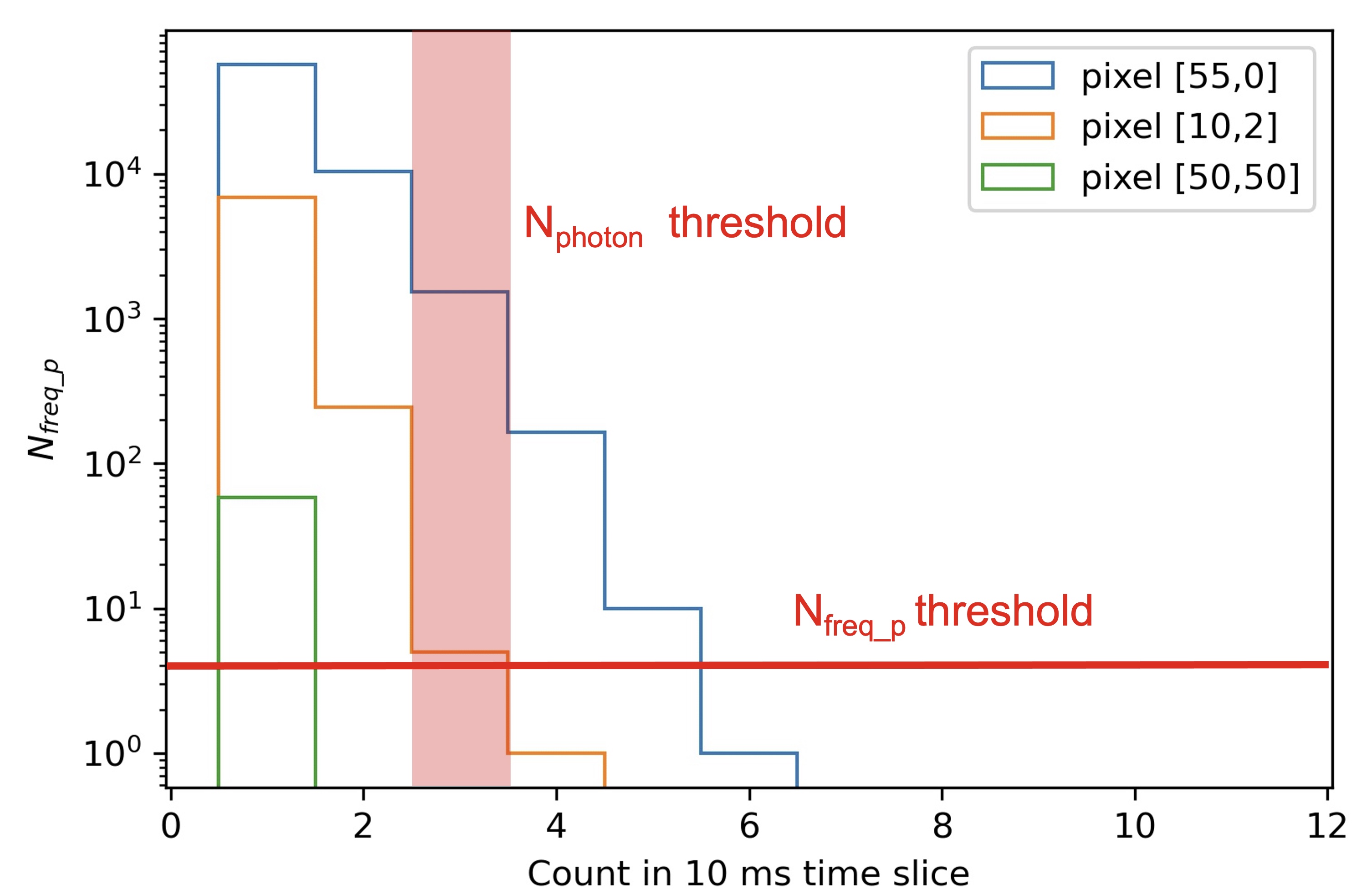}
    \caption{Example of the distribution selection with three pixels: Pixel [55, 0] and pixels [10, 2] are identified as noisy pixels because their $N_{photon}$ and $N_{freq\_p}$ is higher than the selection threshold. The red line and red bins are the thresholds for selecting 2.5\% noisy pixels. }

    \label{fig:distribution_selection}
\end{figure}

\end{enumerate}

\begin{figure}[!h]
    \centering
    \subfigure[]{\label{pixelsSlection}\includegraphics[width=0.49\textwidth]{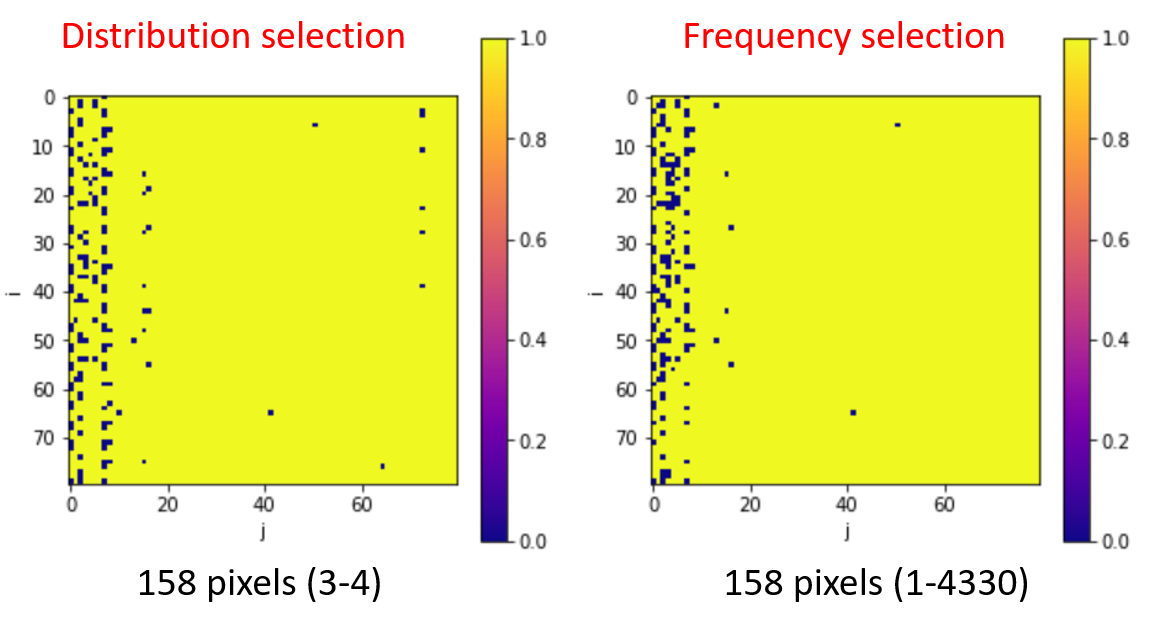}}
    \subfigure[]{\label{pixelsSlectionResult}\includegraphics[width=0.4\textwidth]{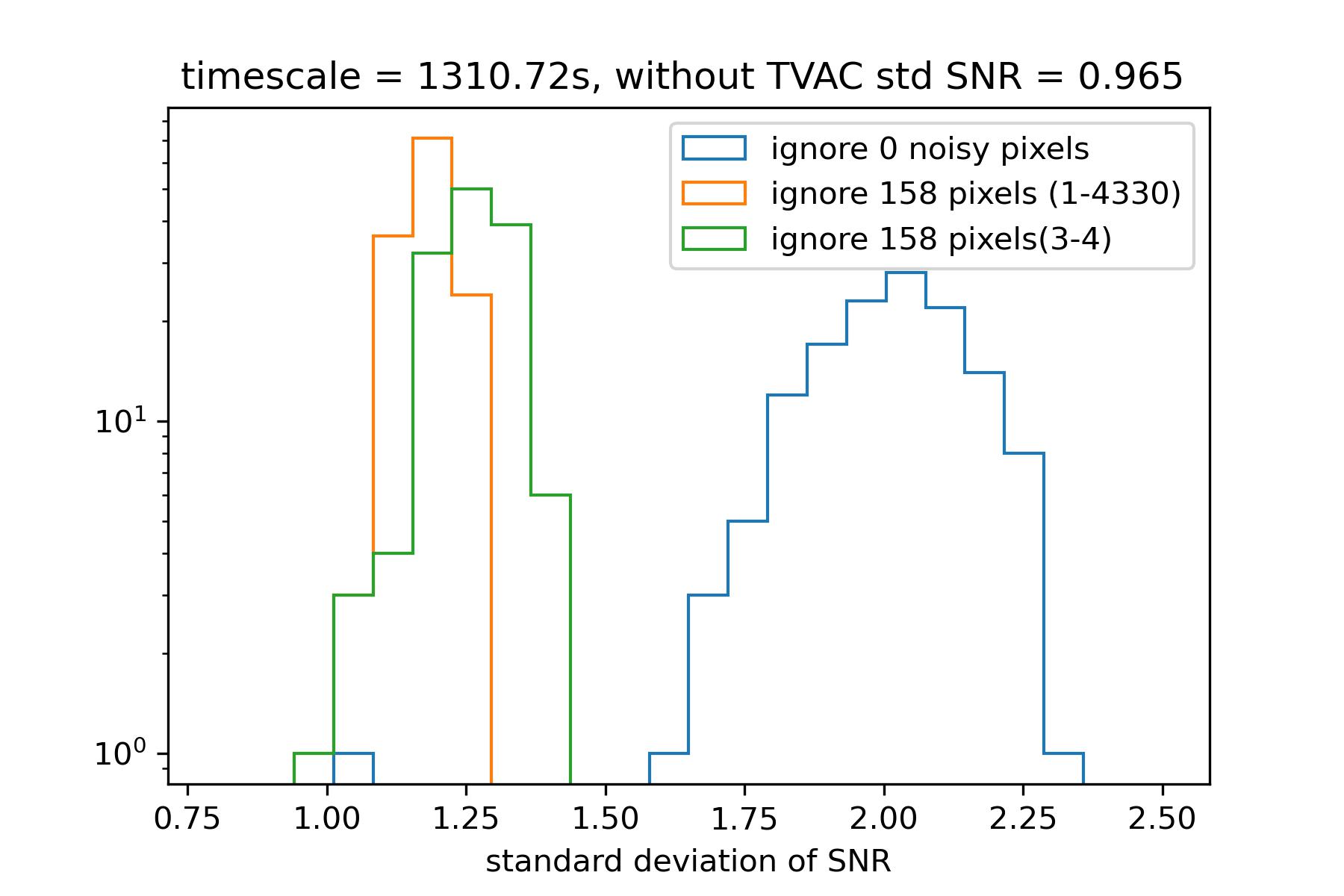}}
        
    \caption{Selected noisy pixel tables and the simulation result after ignoring the selected noisy pixels. 
    (a) Example of 158 heat-pipe noise pixels (black pixels, 2.5\% of the total number) selected through the two selection methods (distribution and frequency selection). 
    (b) Histogram of the $\sigma_{S/N}$ in the 4--8 keV band for the 1310 s timescale after using the different selection methods. The blue histogram shows the result without correction for the heat-pipe noise.
    (the mean value is about 2, which means that the trigger threshold would have to be increased by a factor of 2 to avoid false triggers). 
    The frequency selection method is shown in green and the distribution selection method is shown in orange (after those corrections, the trigger thresholds need to be increased only slightly compared to the ideal case without heat-pipe noise).
    }

    \label{fig:CEpixelsSelection}
\end{figure}

The number of rejected pixels can be tuned by adjusting the value of the thresholds in both selection methods. By selecting a loss of 2.5$\%$ noisy pixels in the detection plane, we can reduce the threshold increment to about 20$\%$ for a 20 min timescales, as shown in Fig.\ref{fig:CEpixelsSelection}. While our current approach has successfully reduced the impact of noise pixels in our simulation, we recognize that there is still room for improvement. We have explored the possibility of selecting additional noisy pixels, but this may result in a trade-off with the loss of sensitivity.

In order to quantitatively compare the effectiveness of the two different selection methods,
for each given timescale, all simulated time intervals with heat-pipe noise were processed, and their $\sigma_{S/N}$ were computed. 
We then computed the average of the 100 highest $\sigma_{S/N}$, and we compared this value with the ideal value $\sigma_{S/N}=1$. 
This value corresponds to the increase in the trigger threshold that is required to avoid false triggers due to the heat-pipe noise. We found that the timescales with the largest threshold increment for CRT and IMT are 10 ms and 20 min, respectively. Therefore, we used these two timescales to compare the effectiveness of the selection methods. The results are shown in Table \ref{table:increment of S/N}.

When we ignored the 158 noisy pixels (about 2.5 $\%$ of total) in the 4--8 keV band, compared to the ideal value $\sigma_{S/N}=1$, the top 100 most strongly impacted cases show an increment of the trigger threshold equal to 18$\%$ for the distribution selection method and 23$\%$ for the frequency selection method in the 10 ms timescale simulation. In contrast, in the 20 min simulation, the increment is 28$\%$ for the distribution selection method and 20$\%$ for the frequency selection method. This suggests that the distribution selection method is better adapted to short-timescale observations, while the frequency selection method is better adapted for long-time observations. This conclusion is also applicable when a higher percentage of noisy pixels is lost.

\begin{table*}[!h]
\centering
\caption{Mean increase in the $\sigma_{S/N}$ (compared to an ideal value of 1), derived from 100 cases of each simulated timescale with the highest $\sigma_{S/N}$ (shortest and longest timescales shown).}
\begin{tabular}{cc|ccccc|}
\cline{3-7}
\multicolumn{1}{l}{}                                          & \multicolumn{1}{l|}{} & \multicolumn{5}{c|}{Pixels number}                                                                                                                                     \\ \hline
\multicolumn{1}{|c|}{Noisy pixels selection method}           & Timescales            & \multicolumn{1}{c|}{0}     & \multicolumn{1}{c|}{158 ($\sim2.5\%$)} & \multicolumn{1}{c|}{322 ($\sim5\%$)} & \multicolumn{1}{c|}{472 ($\sim7.5\%$)} & 645 ($\sim10\%$) \\ \hline
\multicolumn{1}{|c|}{\multirow{2}{*}{Distribution selection}} & 0.01 s                & \multicolumn{1}{c|}{39\%}  & \multicolumn{1}{c|}{18\%}              & \multicolumn{1}{c|}{16\%}            & \multicolumn{1}{c|}{11\%}              & 8\%              \\ \cline{2-7} 
\multicolumn{1}{|c|}{}                                        & 1310.72 s             & \multicolumn{1}{c|}{107\%} & \multicolumn{1}{c|}{28\%}              & \multicolumn{1}{c|}{19\%}            & \multicolumn{1}{c|}{18\%}              & 18\%             \\ \hline
\multicolumn{1}{|c|}{\multirow{2}{*}{Frequency selection}}    & 0.01 s                & \multicolumn{1}{c|}{39\%}  & \multicolumn{1}{c|}{23\%}              & \multicolumn{1}{c|}{20\%}            & \multicolumn{1}{c|}{18\%}              & 14\%             \\ \cline{2-7} 
\multicolumn{1}{|c|}{}                                        & 1310.72 s             & \multicolumn{1}{c|}{107\%} & \multicolumn{1}{c|}{20\%}              & \multicolumn{1}{c|}{18\%}            & \multicolumn{1}{c|}{17\%}              & 17\%             \\ \hline
\end{tabular}
\label{table:increment of S/N}
\end{table*}

\begin{figure}[!h]
\centering
\subfigure[]{\label{fig: triggerRate4--8 a}\includegraphics[width=.49\textwidth]{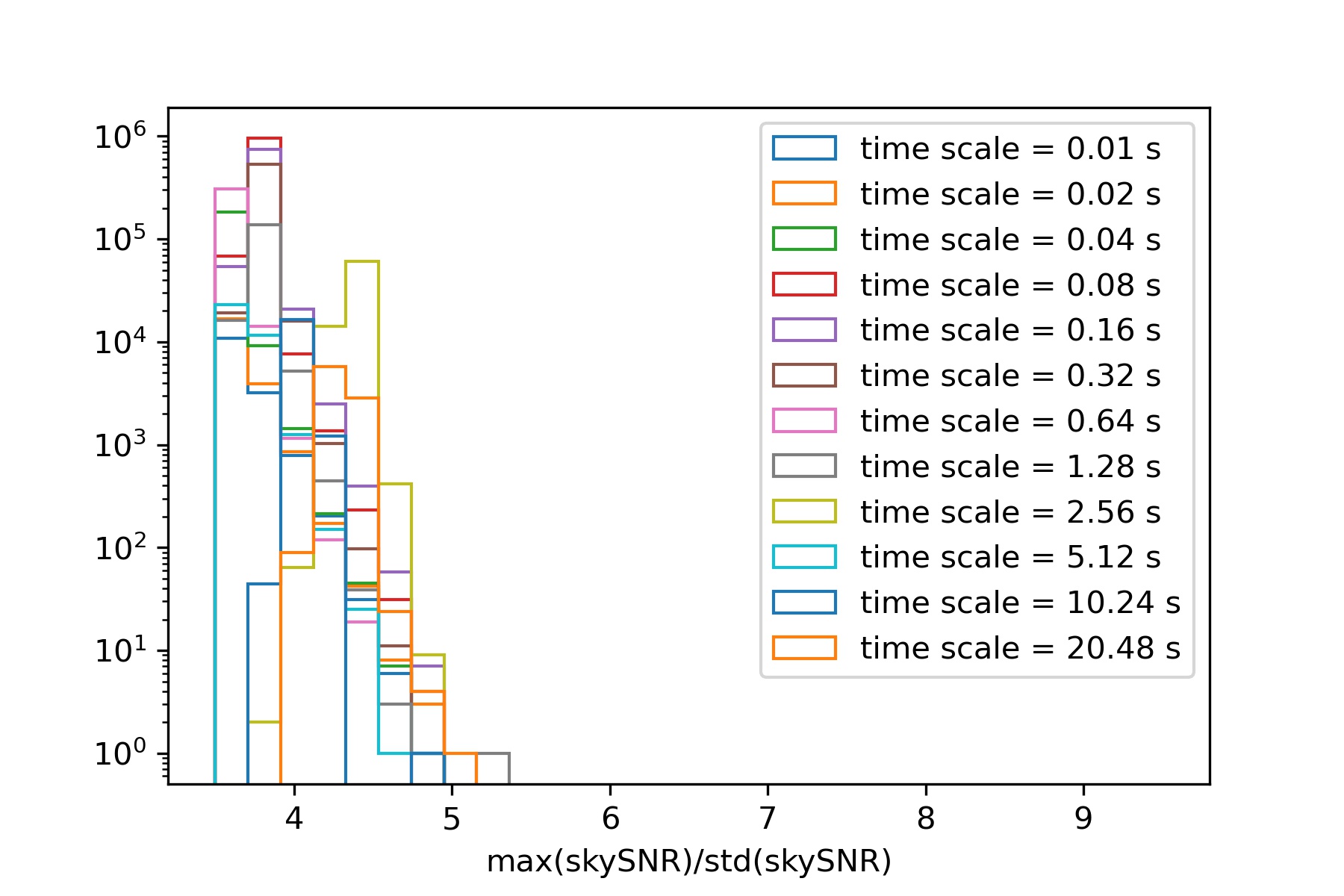}}
\subfigure[]{\label{fig: triggerRate4--8 b}\includegraphics[width=.49\textwidth]{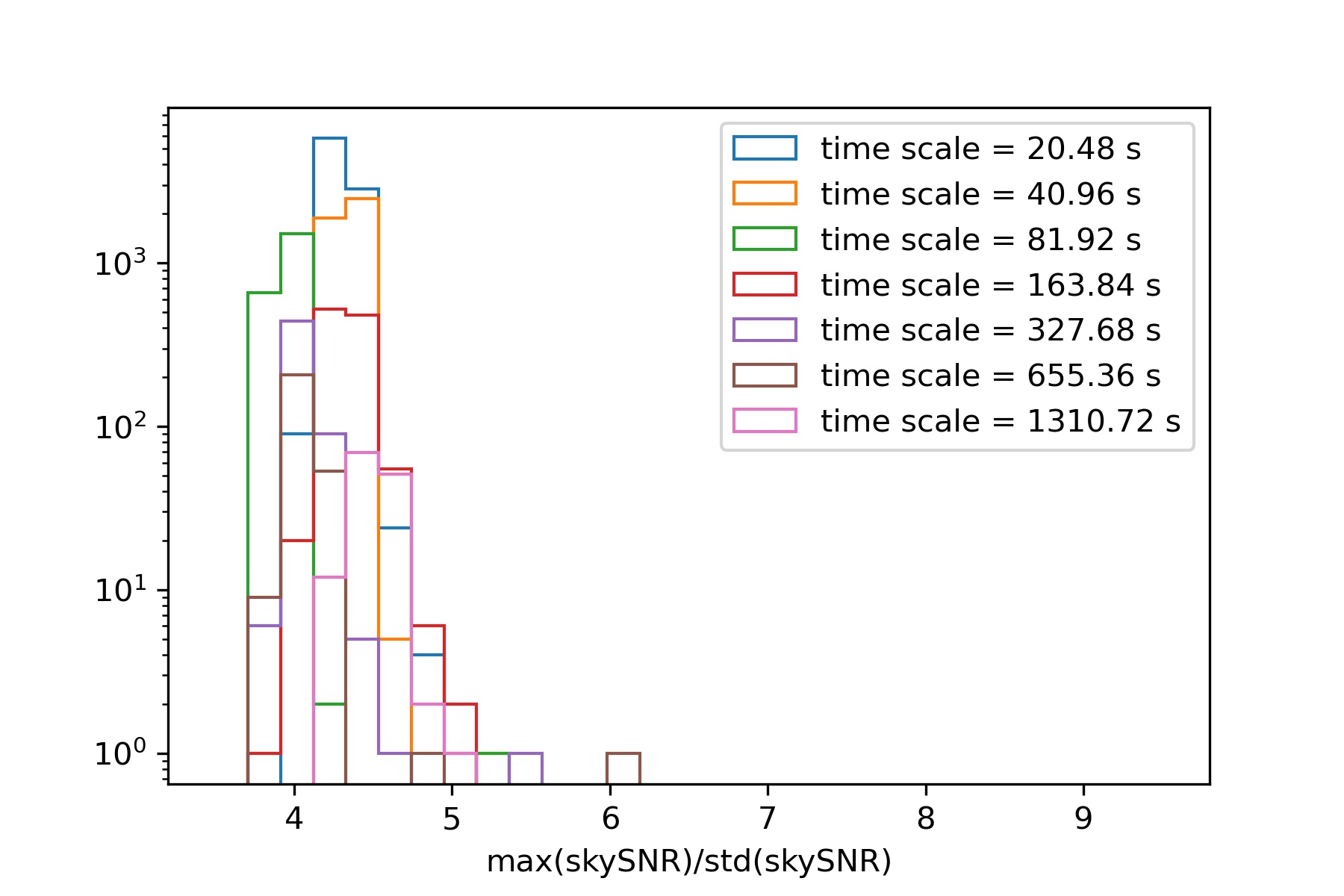}}

\caption{After removal of 158 heat-pipe noise pixels, panels (a) and (b) show the histograms of the $max(S/N)/\sigma_{S/N}$ for different timescales in the 4--8 keV band. Panel (a) is used for the CRT and panel (b) shows the IMT. In each deconvolved S/N sky map, we selected the pixel with the maximum S/N value and computed the $\sigma_{S/N}$ of the sky map. For a given timescale, the histogram shows the distribution of the ratio $max(S/N)/\sigma_{S/N}$ for all-sky maps of this timescale. 
If $max(S/N)/\sigma_{S/N} > 6.5$, it means that there is a false trigger. 
Finally, no false trigger appears on any timescale after our heat-pipe noise removal procedure. }
\label{fig: triggerRateIn160Pixels4To120keV}
\end{figure}

After removing the data from the 2.5$\%$ selected pixels using either the frequency or distribution method, no false triggers were left for all timescales in the 4--8 keV and 4--120 keV energy bands. Fig.\ref{fig: triggerRateIn160Pixels4To120keV} shows the distribution of max(S/N)/$\sigma_{S/N}$ in the 4--8 keV band after $\sim$ 2.5 $\%$ (158) selected noisy pixels were removed based on the frequency selection method. None of the simulated cases in the timescales from 10 ms to 20.48 s had a value above 6.5, which is the onboard trigger threshold value for (S/N)/$\sigma_{S/N}$. Similarly, there were no false triggers for timescales from 20.48 s to 20 min. The same situation also applies to the 4--120 keV energy band.

\section{Discussion} \label{Sec:Discussion}

It is very interesting that a gamma-ray telescope with a large field of view extends the detection energy band down to 4 keV with a semiconductor detector camera.
Before ECLAIRs, a similar instrument such as the imaging telescope on board the INTEGRAL satellite (IBIS, \citet{Lebrun2003}) starts detecting photons at 15 keV. The Burst Alert Telescope (BAT) on board the Swift mission \citep{Gehrels2005} also have a detection range beginning at 15 keV. We discuss some of the challenges we have encountered in this newly extended space below.

\subsection{Causes and consequences of the efficiency inhomogeneity}

The pixel efficiency was found to be inhomogeneous below 8 keV from on-ground calibration data. This might be caused by different reasons. The CdTe pixels in the ECLAIRs detection plane were manufactured in two different periods in 2008 and 2016. The detectors with fewer counts in the detection plane mainly come from the 2016 batch. The difference between the two batches may be due to variations in the detector parameters. The CdTe detector bare element is composed of a platinum (Pt) cathode and an indium (In)-titanium (Ti) anode surrounding the CdTe crystal. There may also be a dead layer composed of tellurium dioxide (TeO$_{2}$) at the interface between the cathode and the sensitive bulk of CdTe \citep{Dubos2013}. 
In this case, photons have to pass through the Pt cathode and dead layer prior to depositing their energy in the detector.
The efficiency difference between the two pixel batches changes with energy. This indicates that the difference may come from a variation in the thickness of the Pt and/or TeO$_{2}$ layers between the two batches. This would result in a different absorption intensity. 

By assuming that there is only one material with a different thickness between the two populations of pixels, either Pt or TeO$_{2}$, this difference can be calculated according to \autoref{equ:thickness calculation}.

\begin{equation}
x_2-x_1=\frac{\ln \left(\frac{I_1}{I_2}\right)}{-\mu}
\label{equ:thickness calculation}.
\end{equation}

In the equation, $x_2-x_1$  stands for the thickness difference (Pt or TeO$_{2}$) between the two populations, $I_1$ and ${I_2}$ represent the mean counts of the Gaussian fits of the two populations in Fig. \ref{fig:fig2}, and $\mu$ represents the attenuation factor of the material. The results of the calculations are shown in Table \ref{Tab: thickness difference}.

\begin{table*}[!h]
    \centering
    \caption{Estimated difference in thickness of the Pt or TeO$_{2}$ material. These results are deduced from \autoref{equ:thickness calculation} in a model where the efficiency difference between LEP and HEP population is due to the Pt thickness differences alone or to the TeO$_{2}$ thickness differences alone. }
\begin{tabular}{|c|c|c|c|c|c|c|c|}
\hline $\begin{array}{c}\text { Energy } \\
(\text{keV)}\end{array}$ & Peak1\tablefootmark{a} & Peak2\tablefootmark{a} & $\begin{array}{c}\text { Peak } \\
\text{difference}\end{array}$ & $\mu_{Pt}$ $(/ \mathrm{nm})$ & $\begin{array}{l}\text { Pt thickness}  \\
\text { difference }(\mathrm{nm})\end{array}$ & $\mu_{TeO_2}$ $(/ \mathrm{nm})$ & $\begin{array}{l}\mathrm{TeO}_2 \text { thickness}  \\
\text { difference (nm) }\end{array}$ \\
\hline 4.5 & 0.888 ($\pm$5.41\%) & 1.063 ($\pm$4.63\%) & 0.175$\pm$0.069 & 0.00171 & 105$\pm$42 & 0.000337 & 534$\pm$210 \\
\hline 4.9 & 0.887 ($\pm$5.19\%) & 1.066 ($\pm$4.52\%) & 0.179$\pm$0.067 & 0.00138 & 133$\pm$50 & 0.000367 & 501$\pm$189 \\
\hline 5.4 & 0.911 ($\pm$4.17\%) & 1.052 ($\pm$3.80\%) & 0.141$\pm$0.055 & 0.00109 & 132$\pm$52 & 0.000335 & 430$\pm$168 \\
\hline 5.9 & 0.939 ($\pm$4.37\%) & 1.049 ($\pm$3.33\%) & 0.110$\pm$0.054 & 0.000871 & 127$\pm$63 & 0.00025 & 443$\pm$220 \\
\hline 6.4 & 0.953 ($\pm$3.15\%) & 1.035 ($\pm$2.99\%) & 0.082$\pm$0.043 & 0.000713 & 116$\pm$61 & 0.000211 & 391$\pm$206 \\
\hline
\end{tabular}
\tablefoot{
\tablefoottext{a}{Peak1 and Peak2 correspond to the mean values of the Gaussian fit in Fig. \ref{fig:fig2}.}}
\label{Tab: thickness difference}
\end{table*}

First, we assumed that the efficiency difference between the two populations of pixels is caused by a difference in the Pt layer thickness alone. In this model, the thickness difference should have a mean value $\sim$ 123 nm, which is derived from the calculation result in Table \ref{Tab: thickness difference}. Then, we assumed that the efficiency difference is caused by the TeO$_{2}$ layer alone. In this second model, the thickness difference of TeO$_{2}$ layer should be $\sim$ 460 nm.
In both cases, the consistency of the thickness difference values found at different energies in Table \ref{Tab: thickness difference} supports the hypothesis of a larger thickness of either Pt or TeO$_{2}$ in the 2016 batch.
A more realistic possibility is that the difference in the pixel efficiency between the two populations is the result of the combined effect of different thicknesses of Pt and TeO$_{2}$.

Although we can configure the trigger software to mitigate the impact from the inhomogeneities of the detection plane in the 4--8 keV band, the effective area of the ECLAIRs detection plane is inevitably decreased by approximately 100 cm$^2$ in this energy band. Compared to the ideal detection plane with only HEPs, the HTPs reduce the effective detection area by 6.25$\%$. The LEPs reduce the effective detection area by 9.53$\%$, 9.24$\%$, 6.77$\%$, 6.24$\%$, and 4.21$\%$ at 4.5 keV, 5.0 keV, 5.4 keV, 5.9 keV, and 6.4 keV, respectively.

As a consequence, the detection sensitivity will decrease by a few percent in this energy band, and the detection rate of soft GRBs (e.g., X-ray flashes) might be reduced. More detailed studies need to be conducted in the future to quantify this rate.

\subsection{Reasons for the heat-pipe noise and possible solution during the operation phase}

Based on our thorough investigation of the origin of the heat-pipe noise, we established a link between the presence of the noise and the activation of the heat pipes. 
During the on-ground calibration measures, the detection plane was installed in a vertical configuration, preventing the heat pipes from working in a nominal manner (i.e., under microgravity). This may have played a role in producing this noise. 
We pursued our investigation on the ground model of the detection plane (one-quarter) that will help us understand the behavior of the detection plane when it is in space. 
\cite{Godet2022} made some measures within a TVAC chamber, this time, with the detection plane in an almost horizontal configuration. The result showed that the heat-pipe noise seems to disappear completely.
This indicates that the operation of the heat-pipes may be affected by gravity (in contrast to space, where they are operating in a free-fall environment). Therefore, they may not be working normally in the vertical position. This can be explained by the fact that gravity affects the fluid diffusion inside the heat pipes, generating microvibrations in the evaporator. It is therefore possible that heat-pipe noise may be absent during ECLAIRs operations in space.

In the operation phase, SVOM will follow a roughly antisolar pointing strategy, which induces Earth's presence in the field of view of ECLAIRs.
If the heat-pipe noise occurs during the ECLAIRs operations in space, it will be superimposed on a background dominated by the CXB, which changes with Earth’s occultation. To identify heat-pipe noise pixels, we propose that the data are analyzed that will be obtained during Earth’s occultations because then, no X-ray sources are visible and only a few CXB counts are detected during this time. In this case, no special observation program is required to obtain data like this. With these data, we can identify pixels affected by heat-pipe noise and adjust the weights of the fitting array and deconvolution array from the ground (as described in \ref{subsection: Heat-pipe noise mitigation methods}). Then these arrays will be uploaded and used by the onboard trigger software. This will help to avoid false triggers and mitigate the impact of heat-pipe noise on the trigger threshold during ECLAIRs operations.

\section{Conclusion} \label{Sec:Conclusion}

In 2021, a series of tests were carried out to measure the performance of the ECLAIRs detection plane. We found that the pixel efficiency in the 4--8 keV band is inhomogeneous, and heat-pipe noise exists on two sides of the detection plane.

A simulation was conducted to study the impact of LEPs and HTPs. Ideally, the S/N of the sky maps is normally distributed with $\sigma_{S/N}\sim$ 1 on a 20 min observation timescale. After we introduced the effect of efficiency inhomogeneity, the  $\sigma_{S/N} $ in the 4--8 keV band increased to 5.75 and 1.43 for HTPs and LEPs, respectively. Most of the impact of HTPs can be corrected for by setting their weights to 0 in the background fitting table and in the deconvolution table of the trigger algorithm, which means that these pixels are excluded during the data processing in the trigger. To correct for the impact of LEP, the efficiency correction in the shadowgram before the deconvolution seems to be a good solution.

We have investigated the characteristics of heat-pipe noise and its impact on the detection and studied methods that might mitigate its effects. Pixels affected by heat-pipe noise display a relatively high count rate in the 4--8 keV band compared to those without heat-pipe noise. This effect not only increases the trigger threshold, but also leads to false triggers, thereby reducing the GRB trigger sensitivity, particularly during long-duration observations in the image trigger (IMT).

The introduction of heat-pipe noise counts from TVAC data in 20 min simulations results in an increased trigger threshold of approximately 100$\%$ (6.5  $\sigma_{S/N} $) compared to observations without heat-pipe noise. However, even with this increased threshold, a false-trigger rate of 99.26$\%$ in the 4--8 keV band and 4.44$\%$ in the 4--120 keV band was observed on the longest duration timescale of 20 min.

We developed methods for mitigating the impact of heat-pipe noise by selecting the noisy pixels. By accepting a loss of 2.5$\%$ - 5$\%$ pixels, we can prevent false triggers caused by heat-pipe noise (false-trigger rate = 0$\%$) and reduce the threshold increment to about 20$\%$ on the longest 20 min trigger timescale in the 4--8 keV band.

\begin{acknowledgements}
ECLAIRs is a cooperation between CNES, CEA and CNRS, with CNES acting as the prime contractor.
This work is supported by the Program of China Scholarships Council (grant No. 202006660005).
We would like to thank the ECLAIRs teams at CNES and IRAP for making the calibration data available to us.
We also acknowledge Olympe Léchine who worked on the data processing at the beginning of this study.
\end{acknowledgements}

\bibliographystyle{aa}
\bibliography{references}  

\end{document}